\documentclass{aastex}
\usepackage{emulateapj5}

\slugcomment{Astrophysical Journal, January 2002}
\shortauthors{Fitzpatrick et al.}
\shorttitle{Distance to the LMC}

\begin{document}

\title{The Fundamental Properties and Distances of LMC Eclipsing Binaries II. HV~982{\footnote{Based on observations with the NASA/ESA Hubble Space Telescope, obtained at the Space Telescope Science Institute, which is operated by the Association of Universities for Research in Astronomy, Inc. under NASA contract No. NAS5-26555.}}}

\author{E.L.~Fitzpatrick,~I.~Ribas\altaffilmark{2},~E.F.~Guinan\altaffilmark{2},~L.E.~DeWarf\altaffilmark{2},~F.P.~Maloney}
\affil{Department of Astronomy \& Astrophysics, Villanova University, Villanova, PA 19085}

\author{D.~Massa}
\affil{Emergent IT}

\altaffiltext{2}{Visiting Astronomer, Cerro Tololo Inter-American Observatory, National Optical Astronomy Observatories, which is operated by the Association of Universities for Research in Astronomy, Inc. (AURA) under cooperative agreement with the National Science Foundation.}

\begin{abstract}

We have determined the distance to a second eclipsing binary system
(EB) in the Large Magellanic Cloud, HV~982 ($\sim$B1 IV-V + $\sim$B1
IV-V).   The measurement of the distance --- among other properties of
the system --- is based on optical photometry and spectroscopy and
space-based UV/optical spectrophotometry.  The analysis combines the
``classical'' EB study of light and radial velocity curves, which
yields the stellar masses and radii, with a new analysis of the
observed energy distribution, which yields the effective temperature,
metallicity, and reddening of the system plus the distance
``attenuation factor'', essentially $({\rm radius/distance})^2$.
Combining the results gives the distance to HV~982, which is $50.2 \pm
1.2$ kpc.
 
This distance determination consists of a detailed study of
well-understood objects (B stars) in a well-understood evolutionary
phase (core H burning).   The results are entirely consistent with ---
but do not depend on --- stellar evolution calculations.  There are no
``zeropoint'' uncertainties as, for example, with the use of Cepheid
variables.  Neither is the result subject to sampling biases, as may
affect techniques which utilize whole stellar populations, such as red
giant branch stars.  Moreover, the analysis is insensitive to stellar
metallicity  (although the metallicity of the stars is explicitly
determined) and the effects of interstellar extinction are determined
for each object studied.

After correcting for the location of HV~982, we find an implied
distance to the optical center of the LMC's bar of $d_{\rm LMC} =
50.7\pm1.2$ kpc.  This result differs by nearly 5 kpc from our earlier
result for the EB HV~2274, which implies a bar distance of 45.9 kpc.
These results may reflect either marginally compatible measures of a
unique LMC distance or, alternatively, suggest a significant depth to
the stellar distribution in the LMC.  Some evidence for this latter
hypothesis is discussed.
\end{abstract}


\section{Introduction}

The distance to the Large Magellanic Cloud (LMC) is a key factor in
determining the size scale of the Universe.  Indeed, the uncertainty in
the length of this ``cosmic meter stick'' is responsible for much of
the current uncertainty in the value of the Hubble constant, as noted
by Mould et al. (2000).  The LMC distance is as controversial as it is
important.  Existing determinations span a wide range (see Fig. 1 of
Mould et al.) and are often grouped into the ``long'' distance scale
results ($d > 50$ kpc and ($V_0-M_V) \simeq 18.7$ mag) and the
``short'' distance scale results ($d < 50$ kpc and ($V_0-M_V) \simeq
18.3$ mag).  In some cases, the same technique (e.g., Cepheids) can
support both the long and short scales, depending on the assumptions
and details of the analysis.  Reviews of LMC distance determinations
can be found in Westerlund (1997) and Cole (1998), and numerous new
papers have appeared in the last several years, indicating the high
activity level and interest in the field.

Recently, we showed that well-detached main sequence B-type eclipsing
binary (EB) systems are ideal standard candles and have the
potential to resolve the LMC distance controversy (Guinan et al. 1998a;
hereafter Paper I).  The advantages of using EBs are numerous.  First,
an accurate distance can be determined for each individual system ---
and there are many systems.  This is in contrast to techniques that
utilize, for example, Cepheids or red giant stars, where entire
populations are used to derive a single distance estimate.  It also
contrasts with analyses of SN 1987A, which have the potential to yield a
precise distance but for which there is only one object to study.  The
EBs can provide not only the mean LMC distance, but also can be used to
probe the structure of the LMC.  Second, the analysis involves
well-understood objects in a well-understood phase of stellar evolution
(core hydrogen burning) and the results for each object can be verified
independently by --- but do not depend on --- stellar evolution
theory.  Third, the analysis is robust and not subject to any zeropoint
uncertainties, nor are there any adjustable parameters.   For example,
the results are extremely insensitive to stellar metallicity, although
the metallicity of the individual EBs is explicitly determined and
incorporated in the analysis.  Likewise, the determination of
interstellar extinction is an integral part of the analysis of each
object.

The study of the EB system HV~2274 in Paper I yielded a distance of
$46.8\pm1.6$ kpc and stellar properties consistent with stellar
evolution theory (Ribas et al. 2000a).  Correcting for the position of
HV~2274 relative to the LMC center yielded a LMC distance of
$45.7\pm1.6$ kpc corresponding to $(V_0-M_V) = 18.30\pm0.07$ mag.  This
result argues strongly in favor of the ``short distance'' to the LMC.
Since Paper I, several partial-reanalyses of the HV~2274 system have
appeared in the literature (Udalski et al. 1998; Nelson et al. 2000;
Groenewegen \& Salaris 2001).  These authors advocate various
adjustments in the results from Paper I which yield LMC distance moduli
in the range 18.22--18.42 mag.  These adjustments all stem from
complications arising from the incorporation of optical photometry in
the analysis.  We will return to this issue later in this paper.
 
The main goal of this paper is to apply our analysis to a second LMC EB
system, HV~982, and derive its stellar properties and distance.  This
EB, with $V \simeq 14.6$, consists of two mildly evolved main sequence
B stars each corresponding to spectral class $\sim$B1 IV--V.  A key
difference between this analysis and that presented in Paper I for
HV~2274 is that we have here obtained space-based spectrophotometric
measurements extending from 1150 \AA\/ to 7500 \AA, eliminating the
reliance on optical photometry.  These new data preclude the
ambiguities which plague the HV~2274 result and allow us to realize the
full potential of our analysis technique.  In \S 2, we describe the
data included in this study.  In \S 3, the analysis --- which
incorporates the light curve, radial velocity curve, and spectral
energy distribution of the system  --- is discussed.   Some aspects of
our results relating to the interstellar medium towards HV~982 are
described in \S4, including an indication of the relative location of
HV~982 within the LMC.  The general stellar properties of the HV~982
system and their consistency with stellar evolution theory are
described in \S 5.  In \S 6, we show how the distance to the system is
derived from our analysis, and compare this result with a reanalysis of
the HV~2274 data.  We discuss the distance to the LMC and summarize our
conclusions from study of two binary systems in \S 7.

\section{The Data}

Three distinct datasets are required to carry out our analyses of the
LMC EB systems: high-resolution spectroscopy (yielding radial velocity
curves), precise differential photometry (yielding light curves), and
multiwavelength spectrophotometry (yielding temperature and reddening
information).  Each of these three is described briefly below.

Note that in this paper, the primary ("P") and secondary ("S")
components of the HV~982 system are defined photometrically and refer
to the hotter and cooler components, respectively.  As we will show,
the primary star is the less luminous and less massive of the pair.

\subsection{Optical Spectroscopy}

Radial velocity curves for HV 982, and a number of other LMC EBs, were
derived from optical echelle spectra obtained by us during 6-night and
8-night observing runs in January and December 2000, respectively, with
the Blanco 4-m telescope at Cerro Tololo Inter-American Observatory in
Chile. The seeing conditions during the two runs ranged between 0.7 and
1.8 arcsec.  We secured eighteen spectra of HV~982 -- near orbital
quadratures -- covering the wavelength range 3600--5500~\AA, with a
spectral resolution of $\lambda/\Delta\lambda \simeq 22000$, and a S/N
of $\sim$20:1.  The plate scale of the data is 0.08~\AA~pix$^{-1}$
(5.3~km~s$^{-1}$~pix$^{-1}$) and there are 2.6 pixels per resolution
element.  Identical instrumental setups were used for both observing
runs. The exposure time per spectrum was 1800 sec, sufficiently short
to avoid significant radial velocity shifts during the integrations.
All the HV~982 observations were bracketed with ThAr comparison spectra
for proper wavelength calibration. The raw images were reduced using
standard NOAO/IRAF tasks (including bias subtraction, flat field
correction, sky-background subtraction, cosmic ray removal, extraction
of the orders, dispersion correction, merging, and continuum
normalization). Spectra of radial velocity standard stars were acquired
and reduced with the same procedure.

\begin{figure*}
\plotone{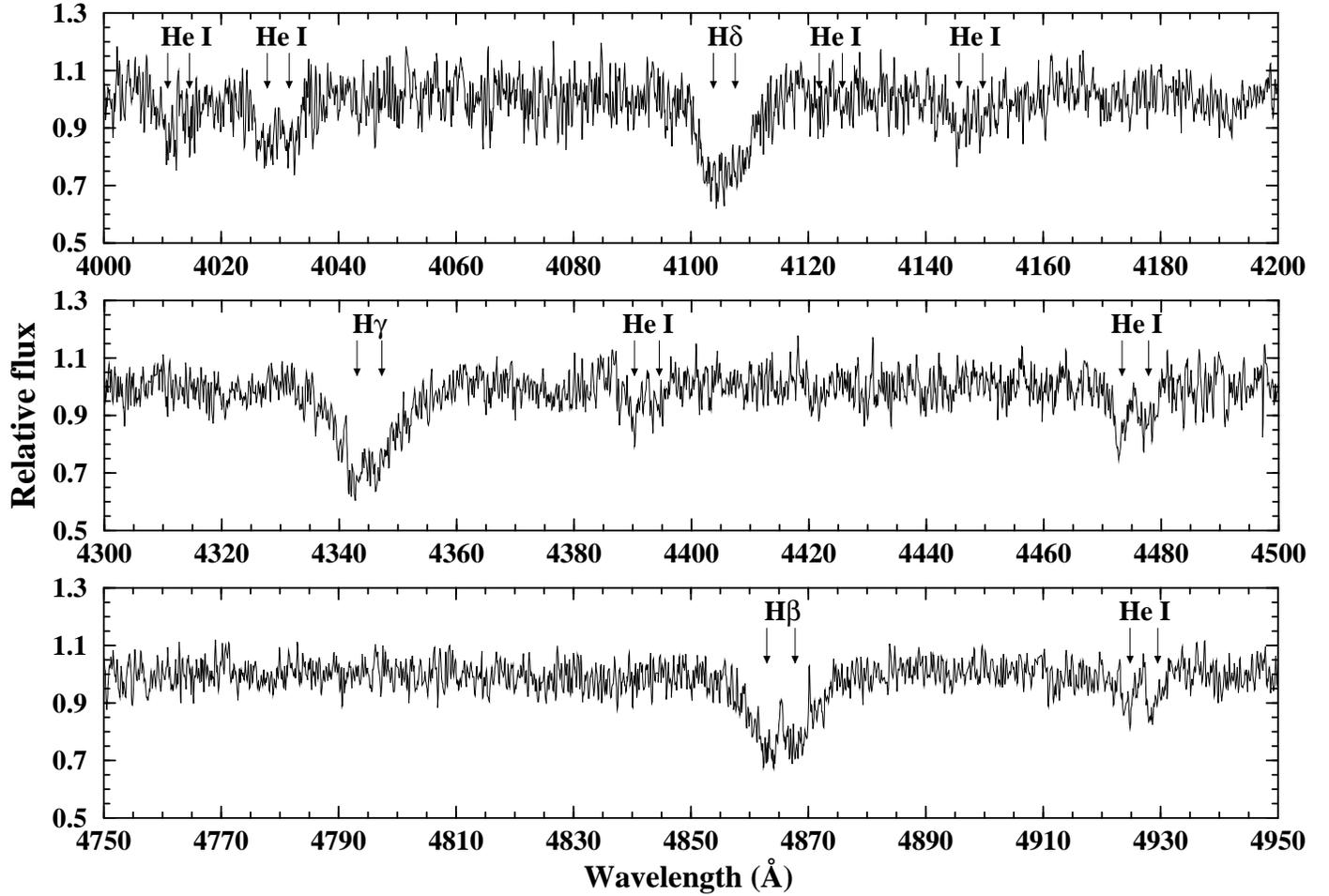}
\caption[f1.eps]{Normalized spectrum of HV~982 near prominent
H~I and He~I lines.  The spectrum was obtained with the Blanco 4-m
telescope at Cerro Tololo Inter-American Observatory on HJD
2451558.7073 at binary phase 0.7252.  The velocity separation between
the primary and secondary stars at this phase is $\Delta$v = 307 km
s$^{-1}$.  This figure demonstrates that the absorption lines from the
two stars are cleanly resolved and, thus, that the radial velocity
measurements will be immune to blending effects.  \notetoeditor{THIS FIGURE IS INTENDED TO SPAN TWO COLUMNS}
\label{figSPEC}}
\end{figure*}

Visual inspection of the HV~982 spectra revealed prominent H Balmer
lines and conspicuous lines of He~{\sc I} (4009, 4026, 4144, 4388,
4471, and 4922~\AA). Various features due to ionized C, Si, and O are
also expected, but with strengths comparable to the noise level in the
individual spectra. For illustration, three 200-\AA\ sections of one of
the observed spectra are shown in Figure \ref{figSPEC}. The strongest
He~{\sc I} features and the H~I Balmer lines are labeled, with arrows
marking the expected line positions for the two components of the
system (according to the radial velocity curve solution described in
\S3.2). This spectrum was obtained at orbital phase 0.725 and
illustrates the clean velocity separation of the two components of the
binary.

\subsection{Optical Photometry}

CCD differential photometric observations of HV~982 were reported by
Pritchard et al.  (1998; hereafter P98).  These data were obtained
between 1992 and 1995 with a 1-m telescope at Mt. John University
Observatory (New Zealand).  The resultant light curves in the
Str\"omgren $u$, Johnson $V$ and Cousins $I$ passbands have very good
phase coverage, with 132, 565, and 205 measurements, respectively.
More sparsely-covered light curves were obtained in the Str\"omgren
$vby$ passbands, with 48, 45, and 44 measurements for $v$, $b$, and
$y$, respectively. The precision of the individual differential
photometric measurements is $0.010-0.015$~mag.

\subsection{UV/Optical Spectrophotometry}
\subsubsection{FOS Data}

We obtained spectrophotometric observations of HV~982 at UV and optical
wavelengths with the Faint Object Spectrograph (FOS) aboard the {\it
Hubble Space Telescope} on 31 January 1997, at binary phase 0.533.
Data were obtained in four wavelength regions, using the G130H, G190H,
G270H, and G400H observing modes of the FOS with the 3.7\arcsec x
1.3\arcsec\/ aperture, yielding a spectral resolution of
$\lambda/\Delta\lambda \simeq 1300$.  The dataset names are Y3FU5503T,
Y3FU5506T, Y3FU5505T, and Y3FU5504T, respectively.  The data were
processed and calibrated using the standard pipeline processing
software for the FOS.  The four individual observations were merged to
form a single spectrum which covers the range 1145 \AA\/ to 4790 \AA.

\subsubsection{STIS Data}

Additional {\it HST} observations of HV~982 were obtained on 22 April
2001, at binary phase 0.621, using the Space Telescope Imaging
Spectrograph (STIS).  Data were obtained in the G430L and G750L
observing modes with the 52\arcsec\/ x 0.5\arcsec\/ aperture, yielding a
spectral resolution of $\lambda/\Delta\lambda \simeq 750$. The dataset
names are O665A8030 and O665A8040. The data were processed and
calibrated using the standard pipeline processing software.  Cosmic ray
blemishes were cleaned ``by hand'' and the spectra were trimmed to the
regions 3510--5690 \AA\/ and 5410--7490 \AA\/ for the G430L and G750L
data, respectively.  Because of concerns about photometric stability
the two STIS spectra were not merged (see \S 3.3.3).

\section{The Analysis}

Our study of HV~982 depends on three separate but interdependent
analyses.  These involve the radial velocity curve, the light curve,
and the observed spectral energy distribution (SED).  The combined results
provide essentially a complete description of the gross physical
properties of the HV~982 system and a precise measurement of its
distance.  Each of the three analyses is described below.  
  
\subsection{The Radial Velocity Curve}
\subsubsection{Measurement of the Radial Velocities}

To determine the radial velocities of the HV~982 components, we
restricted our attention to the 4000--5000~\AA\/ wavelength region of
the spectra described in \S 2.1.  Data at higher and lower wavelengths
were badly contaminated with H Balmer lines, or had very poor
S/N, or both. As is well known, the broad H Balmer lines are generally
not suitable for radial velocity work because of blending effects,
which may lead to systematic underestimation of radial velocity
amplitudes (see, e.g., Andersen 1975). In the 4000--5000 \AA\/ range,
the H$\beta$, H$\gamma$ and H$\delta$ lines were masked by setting the
normalized flux to unity in a window around their central wavelength.

Our initial approach for measuring radial velocities was to use the
cross-correlation technique, with a very high S/N ($\sim$250) spectrum
of HR~1443 ($\delta$~Cae, B2~IV-V, $v \sin i=36$~km~s$^{-1}$) as the
velocity template. Two clean cross-correlation function peaks (one per
stellar component) were clearly visible for all the object spectra as
might be anticipated from Figure \ref{figSPEC}. This allowed us to
determine individual velocities with accuracies of 10-15~km~s$^{-1}$. A
number of tests, however, indicated that the resulting velocities were
moderately dependent on the filtering parameters used. This phenomenon
adds a component of subjectivity to the measured radial velocities, and
prompted us to move to an alternate, and ultimately superior,
technique.

``Spectral disentangling'' is an improvement over classical
cross-correlation because it essentially uses information from the
entire spectral dataset to derive the individual radial velocities.
The basic idea of the technique is very simple: an individual
(observed) double-lined spectrum is assumed to be a linear combination
of two single-lined spectra (one per component) at different relative
velocities (determined by the orbital phase at the time of
observation). The goal is to retrieve the two single-line spectra and
the set of relative velocities for each observed spectrum by
considering the whole dataset simultaneously.  The numerical
implementation is an inversion algorithm of an over-determined system
of linear equations.  In contrast with cross-correlation, this
technique eliminates the need for a spectral template, but it does
require a homogeneous dataset of spectra taken at a variety of orbital
phases.

The practical implementation of the disentangling method has been
carried out using two independent approaches: Simon \& Sturm (1994)
based their algorithm on a singular value decomposition, and Hadrava
(1995, 1997) employed a Fourier transform. In principle, the two
implementations are equally valid and we decided to adopt the Fourier
disentangling code {\sc korel} developed by Hadrava\footnote{Available
from the WWW at http://sunkl.asu.cas.cz/\~{}had/korel.html}. We have
made a number of modifications to the original {\sc fortran} source,
the most significant of which is an increase of the maximum number of
radial velocity bins.

To enhance the performance of the disentangling algorithm, we provided
{\sc korel} with orbital information (period, time of periastron
passage, eccentricity, longitude of periastron) so a reasonable set of
starting values for the velocity semiamplitudes could be computed
(Hadrava 2001, priv. comm.).  Several {\sc korel} runs from different
initial conditions were performed to ensure the uniqueness of the
solution.
 
The final heliocentric radial velocities derived from all the CTIO
spectra using the procedure outlined above are listed in Table
\ref{tabRV} (``RV$_P$'' and ``RV$_S$''), along with the date of
observation and the corresponding phase. The individual errors of the
velocities are not provided by {\sc korel} and a reliable estimation is
not straightforward. This issue is addressed in more detail in \S3.2.

\begin{figure*}
\plotone{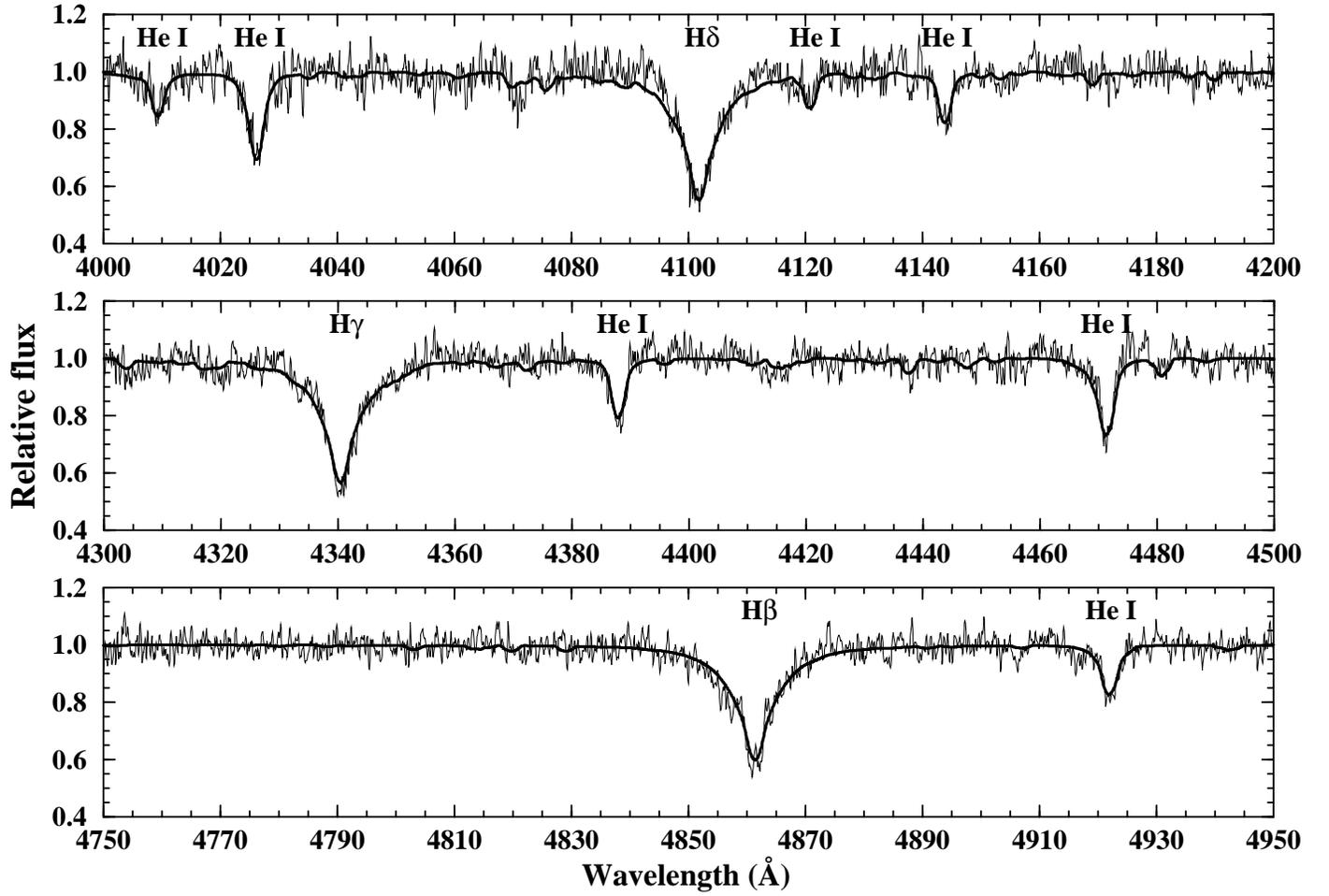}
\caption[f2.eps]{Normalized, ``disentangled'' optical spectrum
of the secondary star in the HV~982 system, compared with a synthetic
spectrum corresponding to $T_{eff} = 23600$ K, $\log g = 3.72$, and $v
\sin i = 106$ km s$^{-1}$.  Prominent H I and He I lines are marked.
The temperature and gravity are determined from our overall analysis;
$v \sin i$ is determined from comparison of the observed spectrum with
a grid of synthetic spectra computed with $v \sin i$ values ranging
from 20 km s$^{-1}$ to 160 km s$^{-1}$.  The synthetic spectra were
produced using R.L. Kurucz's {\it ATLAS9} atmosphere models and I.
Hubeny's spectral synthesis program {\it SYNSPEC}. \notetoeditor{THIS
FIGURE IS INTENDED TO SPAN TWO COLUMNS}
\label{figSECON}}
\end{figure*}

As noted above, the individual spectra for the two components are also
products of the analysis.  Since they combine information from all the
data, the quality of these two spectra is significantly improved with
respect to the individual observations.  Figure \ref{figSECON} shows a
comparison of the disentangled spectrum of the secondary star, i.e, the
more massive and luminous component, with a synthetic spectrum computed
using R.L. Kurucz's {\it ATLAS9} atmosphere models, Ivan Hubeny's
spectral synthesis program {\it SYNSPEC}, and the appropriate stellar
properties (i.e., $T_{\rm eff}$ = 23600 K, $\log g$ = 3.72, ${\rm
[Fe/H] = -0.3}$, and $v \sin i = 106$ km s$^{-1}$).  Note that the
products of {\sc korel} are two spectra that have not been corrected
for ``light dilution,'' i.e., the continuum level contains the light
contribution from the two components and thus the absorption lines are
diluted to roughly half their true strengths.   The spectrum in Figure
\ref{figSECON} has had the primary's contribution removed and then been
renormalized. The contribution of the primary was determined using the
line strengths in the synthetic spectrum as a reference.  The
appropriate value of $v \sin i$ for the model was also determined by
comparing the observed and synthetic spectra.  The derivations of the
basic stellar properties which serve as inputs for the synthetic
spectrum calculations are described below in the rest of \S 3 and all
the stellar properties are summarized in \S 5.  The significance of the
derived $v \sin i$ value is also discussed in \S 5.

As can be seen in Figure \ref{figSECON}, the agreement between the
disentangled and synthetic spectra is excellent, including not only the
depths of all the features but also the profile shapes of both the He I
and H I Balmer lines.  (Recall that the Balmer lines were masked out
when obtaining radial velocities, but a final run of {\sc korel} with
no free parameters was carried out to extract the complete spectrum
that we show in Figure \ref{figSECON}.)  The disentangled spectrum of
the primary star is nearly identical to that of the secondary and shows
similarly good agreement with a synthetic spectrum computed using the
appropriate stellar properties (i.e., $T_{\rm eff}$ = 24200 K, $\log g$
= 3.78, ${\rm[Fe/H] = -0.3}$, and $v \sin i = 85$ km s$^{-1}$).  These
comparisons provide a ``reality check'' for {\sc korel} and, as will be
shown in \S 5, give valuable confirmation for some of the basic results
of our overall analysis.

\subsubsection{Analysis of the Radial Velocity Curve}

The radial velocity data were analyzed using an improved version of the
Wilson-Devinney program (Wilson \& Devinney 1971; hereafter WD) that
includes an atmosphere model routine developed by Milone et al. (1992)
for the computation of the stellar radiative parameters.  The full
analysis can potentially yield determinations of the component velocity
semi-amplitudes $K$ (or, equivalently, the mass ratio $q$), the
systemic velocity $\gamma$, the orbital semi-major axis $a$, the
eccentricity $e$, and the longitude of the periastron $\omega$. In our
particular case, we adopted the very well-determined value of $e$ from
the light curve solution (\S3.2). It is not possible, however, to
utilize the value of $\omega$ yielded by the light curve analysis
without first correcting for the significant apsidal motion of the
system (an increase in $\omega$ of about 12$^{\circ}$ between the epoch
of P98's observations and ours). Instead of using the empirical apsidal
motion rate, we treated $\omega$ as a free parameter and obtained a
best fit for the mean epoch of our data (J2000.5). Thus, solutions were
run in which $q$, $a$, $\gamma$, and $\omega$ were the adjustable
parameters.

\begin{figure*}
\plotone{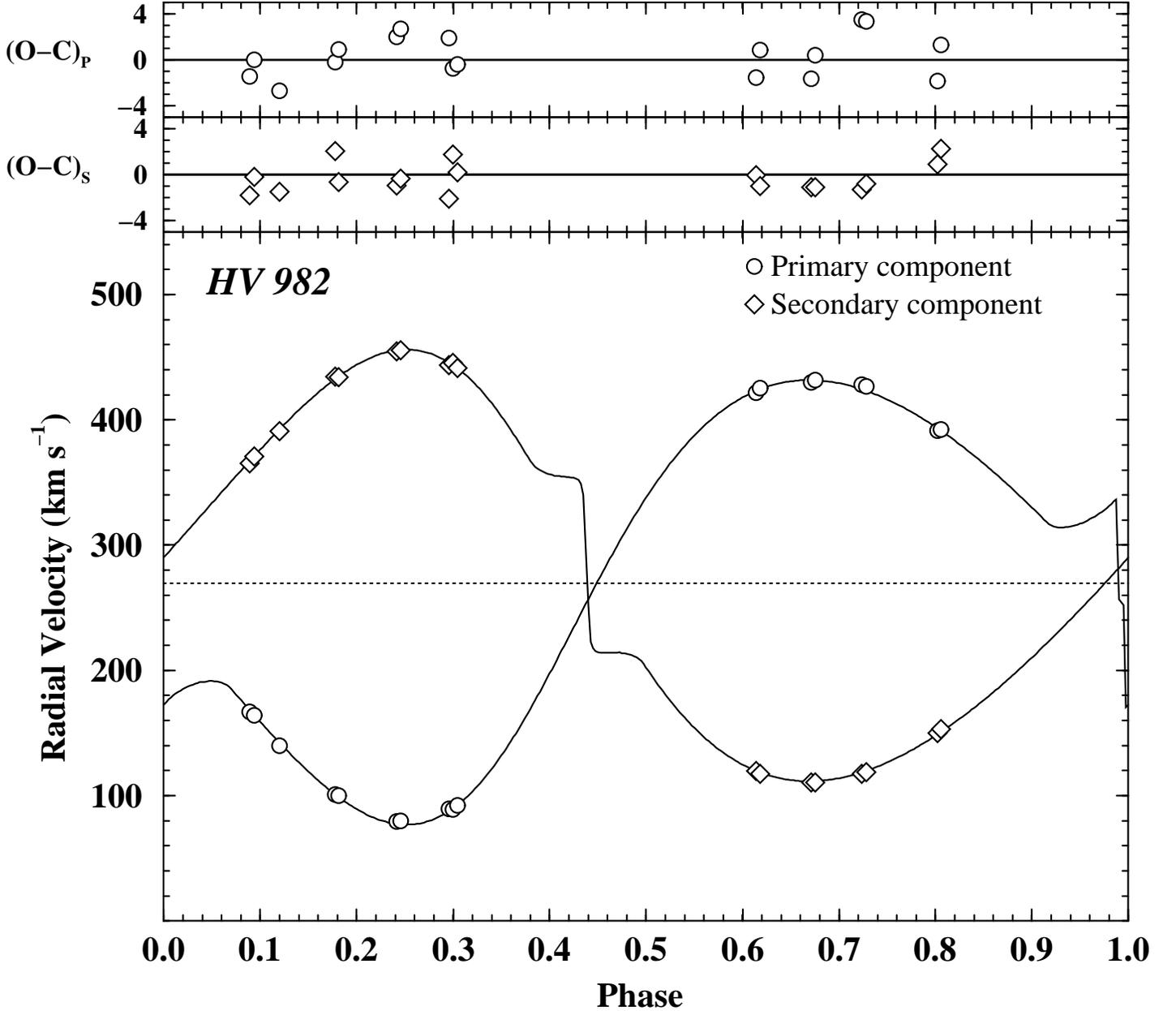}
\caption[f3.eps]{Radial velocity data for HV~982 (see Table 1)
superimposed with best-fitting model. The parameters derived from the
data are listed in Table 2.  Note that the details of the model curve,
including the sharp discontinuity due to the partial eclipse of a
rotating star (the Rossiter Effect), are not a product of the radial
velocity curve analysis. The fit assumed the values of the orbital
eccentricity and inclination found from the light curve.  The
parameters directly determined from the radial velocity curve are the
velocity semi-amplitudes $K$, the systemic velocity $\gamma$, and
longitude of periastron $\omega$.  The residuals to the fit are shown
above the radial velocity curve and indicate r.m.s. uncertainties in
the data of $\sim$1.5~km~s$^{-1}$ for both the primary and secondary
components.
\label{figRV}}
\end{figure*}

Our best fit to the radial velocity curve is shown in Figure
\ref{figRV}. Note that the details of the curve shape (such as its
skewness and the abrupt changes during eclipse) are not a product of
the radial velocity analysis, but rather result from the adopted light
curve solution and from the physical effect of partially-eclipsed
rotating stars (the ``Rossiter Effect''). The fit residuals (indicated
as ``O--C'' in the figure and in the last two columns of Table
\ref{tabRV}) correspond to r.m.s. errors of $\sim$1.5~km~s$^{-1}$. This
small internal error gives an indication of the good quality of the
spectroscopic data and the excellent performance of the disentangling
technique.

The best-fitting parameters to the radial velocity curve are listed in
Table \ref{tabPARMS}. The uncertainties quoted in the table bear some
comment. Indeed, the formal errors derived from the fit to the radial
velocity curve are significantly smaller. For example, the WD program
returns an estimated uncertainty of only $\sim$0.5~km~s$^{-1}$ in the
velocity semiamplitudes.  While formally correct, this uncertainty may
be underestimated because it fails to account for any systematic
effects that may be present in the data.  As discussed by Hensberge,
Pavlovski, \& Verschueren (2000), more realistic estimates of the
uncertainty in the velocity semiamplitudes follow from considering the
scatter of the velocities derived from the analysis of separate
spectral regions.  Thus, we divided our entire spectrum into four
wavelength intervals and analyzed these separately with {\sc korel}.
The standard deviation of the resulting velocities turned out to be
3.2~km~s$^{-1}$. This is most likely an overestimate of the true error
of the velocities because of the spectral coverage being significantly
degraded (and so the number of spectral lines available for radial
velocity determination).  Nonetheless, we conservatively adopted
3.2~km~s$^{-1}$ as the uncertainty of the velocity semiamplitudes and
scaled the rest of the parameter errors listed in Table \ref{tabPARMS}
accordingly.

\subsection{The Light Curve}

P98 ran simultaneous solutions for the 6 available light curves using
the same version of the WD program described above.  The WD program was
run in an iterative mode in order to explore the full-extent of
parameter space and also to make a realistic estimation of the errors.
Furthermore, the authors considered different mass ratios ranging
between 0.9 and 1.1 (since no spectroscopic observations were available
at that time) and found the light curve solution to be completely
insensitive to changes within this range.

As often occurs for eclipsing binary stars in eccentric orbits, several
parameter sets --- four in this case --- were found to yield equally
good fits to the observed light curves. The main distinction among the
possible solutions is that P98's Cases 1 and 2 predict the primary star
to be somewhat smaller and hotter than the secondary (with relative
luminosity in the $V$ and $I$ spectral regions of $[L_{\rm S}/L_{\rm
P}]_{V,I} \simeq 1.1$), while Cases 3 and 4 yield a bigger and cooler
primary ($[L_{\rm S}/L_{\rm P}]_{V,I} \simeq 0.9$). This degeneracy can
be broken only by considering some external source of information,
e.g., a spectroscopically-determined luminosity ratio.  Without access
to such information, P98 were unable to favor any of their four cases.

The disentangled spectra discussed above provide such a spectroscopic
luminosity ratio and allow us to distinguish between P98's two general
scenarios.  As noted in \S3.1, the comparison of the disentangled
spectra with synthetic spectra yields values of $v \sin i$ for both
stars and their light dilution factors, which are related to their
luminosities in the blue ($B$) spectral region.  The ratio of these
dilution factors yields $[L_{\rm S}/L_{\rm P}]_B = 1.15\pm0.05$,
clearly favoring Cases 1 and 2, in which the primary star is smaller
and hotter than the secondary.  Note that these cases are also the
physically preferable ones, since our radial velocity analysis shows
the secondary star to be the more massive member of this
non-interacting main sequence system and, therefore, necessarily the
more luminous star.

\begin{figure*}
\plotone{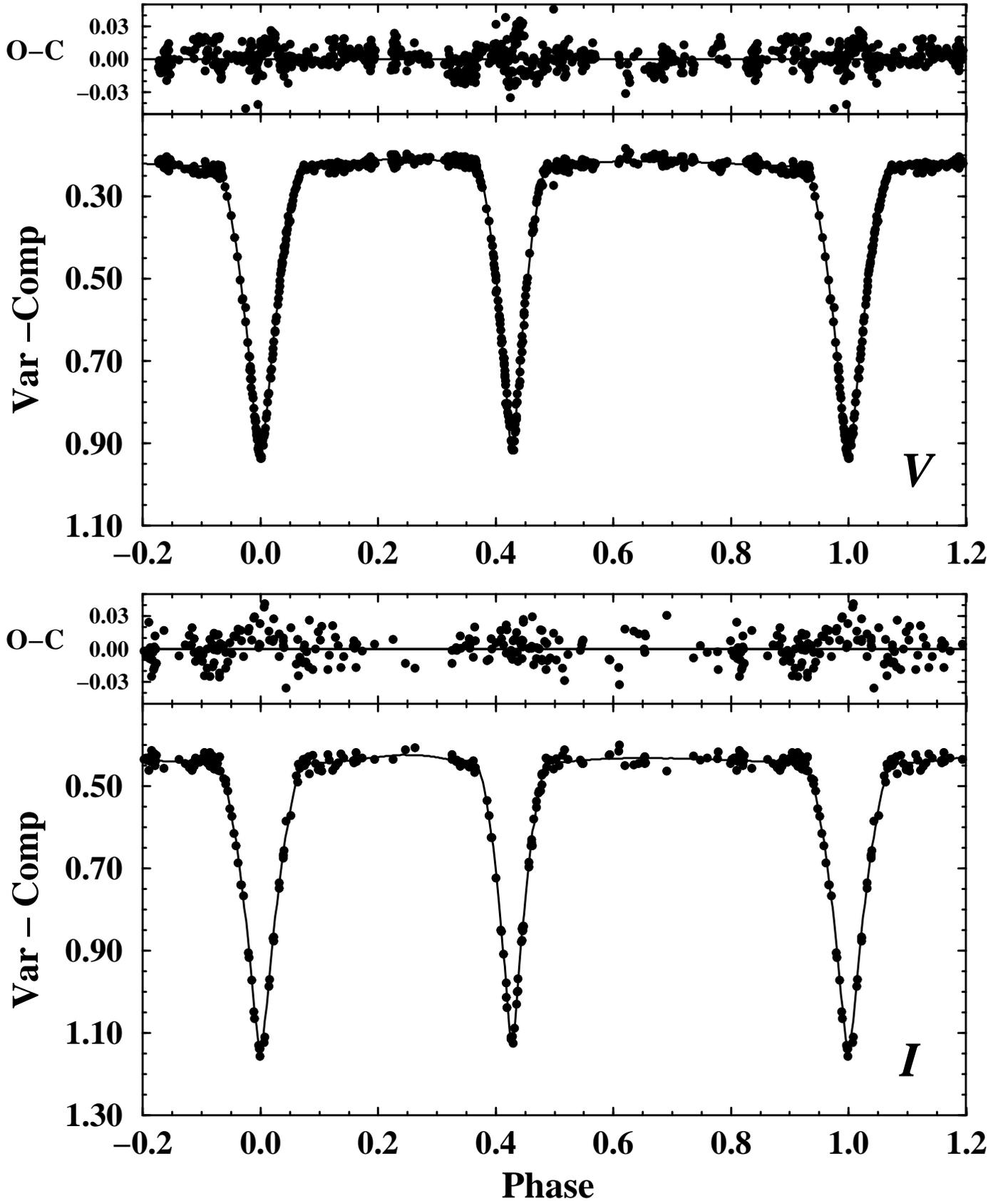}
\caption[f4.eps]{$V$ and $I$ light curves for HV~982 (filled
circles) overplotted with the best fitting model (solid curves). The
residuals to the fits (``O-C'') are shown above each light curve.  The
parameters derived from the fit are listed in Table 2.  The data are
from P98 and the final solution is very similar to P98's Case 2.
\label{figLC}}
\end{figure*}

In an attempt to distinguish between P98's Cases 1 and 2, we redid the
light curve analysis published by P98 using an identical computational
setup (i.e., the iterative WD program).  We applied the WD program to
the observed light curves both individually and as a group,
constraining the spectroscopically-determined mass ratio $q$ and trying
a variety of weighting schemes for the different bandpasses. The tests
revealed P98's Case 1 to be a spurious solution caused by excessive
weight on the Str\"omgren $u$ light curve. This solution never appeared
in the analysis of the well-covered $V$ and $I$ light curves.
Therefore, our results clearly favor P98's Case 2 over the others. In
Figure \ref{figLC} we illustrate this solution to the $V$ and $I$ light
curves.

The final orbital and stellar parameters adopted from the light curve
analysis are listed in Table \ref{tabPARMS}. These were derived from a
simultaneous solution to all the bandpasses, weighted by their
observational errors.  The parameters $r_P$ and $r_S$ represent the
{\em relative} stellar radii, i.e., the physical radii divided by the
orbital semi-major axis $a$. Note that the fractional radii listed are
those corresponding to a sphere with the same volume as the Roche
equipotential (``volume radius'').

\subsection{The UV/Optical Energy Distribution}
\subsubsection{The Fitting Procedure}

The final step in the analysis of HV~982 is modeling the observed shape
of the UV/optical energy distribution.  This procedure is the same as
that used for HV~2274 (see Paper I), and is based on the technique
developed by Fitzpatrick \& Massa (1999; hereafter FM99).

For a binary system, the observed energy distribution
$f_{\lambda\oplus}$ depends on the surface fluxes of the binary's
components and on the attenuating effects of distance and interstellar
extinction.  This relationship can be expressed as:
\begin{eqnarray} \label{basic1}
f_{\lambda\oplus} &=&\left(\frac{R_P}{d} \right)^2 [F_{\lambda}^P + (R_S/R_P)^2 F_{\lambda}^S] \nonumber \\
& &  ~~~~~~~~~~~~~~\times 10^{-0.4 E(B-V) [k(\lambda-V) + R(V)]}
\end{eqnarray}
where $F_{\lambda}^i$ $\{i=P,S\}$ are the surface fluxes of the primary
and secondary stars, the $R_i$ are the absolute radii of the
components, and $d$ is the distance to the binary.  The last term
carries the extinction information, including  $E(B-V)$, the normalized
extinction curve $k(\lambda-V)\equiv E(\lambda-V)/E(B-V)$, and the
ratio of selective-to-total extinction in the $V$ band $R(V) \equiv
A(V)/E(B-V)$.

The analysis consists of a non-linear least squares determination of
the optimal values of all the parameters which contribute to the right
side of equation 1.  We represent the stellar surface fluxes with R. L.
Kurucz's {\it ATLAS9} atmosphere models, which each depend on four
parameters: effective temperature ($T_{\rm eff}$), surface gravity
($\log g$), metallicity ([m/H]), and microturbulence velocity ($\mu$).
In addition, we utilize the six-term parametrization scheme from
Fitzpatrick \& Massa (1990) and the recipe given by Fitzpatrick (1999;
hereafter F99) to construct the wavelength dependent UV-through-IR
extinction curve $k(\lambda-V)$.  Thus, in principle, the problem can
involve solving for two sets of four Kurucz model parameters, the
ratios $(R_P/d)^2$ and $R_S/R_P$, $E(B-V)$, six extinction curve
parameters for $k(\lambda-V)$, and $R(V)$.

For HV~982, several simplifications can be made which reduce the number
of parameters to be determined: (1) the temperature ratio of the two
stars is known from the light curve analysis; (2) the surface gravities
can be determined by combining results from the light and radial velocity
curve analyses and are log g = 3.78 and 3.72 for the primary and
secondary stars, respectively (see \S 5); (3) the values of [m/H] and
$\mu$ can be assumed to be identical for both components; (4) the ratio
$R_S/R_P$ is known; and (5) the standard mean value of $R(V) = 3.1$
found for the Milky Way can reasonably be assumed given the existing
LMC measurements (e.g., Koornneef 1982; Morgan \& Nandy 1982; see \S
4).  With these simplifications in place, we modeled the observed
UV/optical energy distribution of HV~982 solving for the best-fitting
values of $T_{\rm eff}^P$, [m/H]$_{PS}$, $\mu_{PS}$, $(R_P/d)^2$,
$E(B-V)$, and six $k(\lambda-V)$ parameters.

\subsubsection{Preparation of the Data} 
Prior to the fitting procedure, the three spectrophotometric datasets
(one merged FOS spectrum and two STIS spectra) were (1) velocity-shifted
to bring the centroids of the stellar features to rest velocity; (2)
corrected for the presence of a strong interstellar H~I Ly$\alpha$
absorption feature in the FOS spectrum at 1215.7 \AA;  and (3) binned to
match the {\it ATLAS9} wavelength scale.  The Ly$\alpha$ correction was
performed by dividing the spectrum by the intrinsic Ly$\alpha$ profile
corresponding to a total H I column density of ${\rm 1.55\times 10^{21}
cm^{-2}}$, distributed in an LMC component and a Milky Way foreground
component.  The determination of the column densities is discussed in
more detail in \S 4 below.

The binning was accomplished by forming simple unweighted means within
the individual wavelength bins of the {\it ATLAS9} models.  In the
wavelength range relevant to this study, the bin sizes are typically 10
\AA\/ (for $\lambda < 2900$ \AA) and 20 \AA\/ (for $\lambda > 2900$
\AA).  The statistical errors assigned to each bin were computed in the
usual way from the statistical errors of the original data, i.e.,
$\sigma_{bin}^2 = 1/\Sigma(1/\sigma_i^2)$, where the $\sigma_i$ are the
statistical errors of the individual spectrophotometric data points
within the bin.  For all the spectra, these uncertainties typically lie
in the range 0.5\% to 1.5\% of the binned fluxes.

Note that we do not merge the FOS and STIS data into a single spectrum,
but rather perform the fit on the three binned spectra simultaneously
and independently. This is because STIS is less photometrically stable
than FOS and there are likely to be flux zeropoint offsets among the
spectra (see the Instrument Handbooks for FOS and STIS available online
at www.stsci.edu).  We account for this effect in the fitting procedure
by assuming that the FOS data represent the true flux levels and
including two zeropoint corrections (one for each STIS spectrum) to be
determined by the fit.  We later explicitly determine the
uncertainties in the results introduced by zeropoint errors in FOS.

The nominal weighting factor for each bin in the least squares
procedure is given by $w_{bin} = 1/\sigma_{bin}^{2}$.  We exclude a
number of individual bins from the fit (i.e., set the weight to zero)
for the reasons discussed by FM99 (mainly due to the presence of
interstellar gas absorption features).

\subsubsection{Results}
\begin{figure*}
\plotone{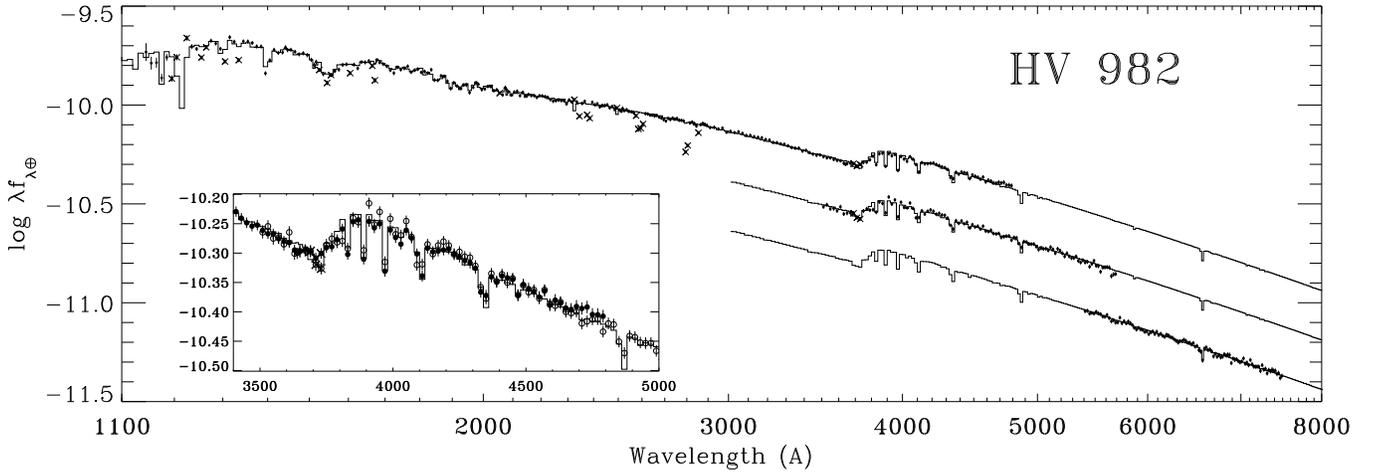}
\caption[f5.eps]{The observed UV/optical energy distribution of
the HV~982 system (small filled circles), superimposed with the
best-fitting model, consisting of a pair of reddened and
distance-attenuated Kurucz {\it ATLAS9} atmosphere models
(histogram-style lines).  Vertical lines through the data points
indicate the $1\sigma$ observational errors.  Crosses denote data
points excluded from the fit, primarily due to contamination by
interstellar absorption lines.  The top spectrum shows the FOS data,
the middle spectrum (shifted by $-0.25$ dex) the STIS/G430L data, and
the lower spectrum (shifted by $-0.5$ dex) the STIS/G750L data.  The
energy distribution fitting procedure was performed simultaneously on
all three datasets.  The inset shows a blowup of the region surrounding
the Balmer Jump which illustrates the overlap between the FOS (solid
circles) and STIS/G430L (open circles) data.  The parameters derived
from the fit to the energy distribution are listed in Tables
\ref{tabPARMS}, \ref{tabSTIS}, and \ref{tabEXT}.  The various
constraints imposed on the fit are discussed in \S3.3.3.
\notetoeditor{THIS FIGURE IS INTENDED TO SPAN TWO COLUMNS}
\label{figSED}}
\end{figure*}

The best-fitting values of the energy distribution parameters and their
1-$\sigma$ uncertainties are listed in Tables \ref{tabPARMS} (stellar
properties), \ref{tabSTIS} (STIS offsets), and \ref{tabEXT} (extinction
curve parameters).  A comparison between the observed spectra and the
best-fitting model is shown in Figure \ref{figSED}.  The three binned
spectra are plotted separately in the figure for clarity (small filled
circles).  The zeropoint offset corrections (see Table \ref{tabSTIS})
were applied to all STIS spectra in Figure \ref{figSED}.  Note that we
show the quantity $\lambda$f$_{\lambda\oplus}$ as the ordinate in
Figure \ref{figSED} (rather than f$_{\lambda\oplus}$) strictly for
plotting purposes, to ``flatten out'' the energy distributions.
 
Our final fit to the HV~982 energy distribution was computed after
adjusting the weights in the fitting procedure to yield a final value
of $\chi^2 = 1$.  Although Figure \ref{figSED} shows that the model
provides an extremely good fit to the data, nevertheless the overall
r.m.s. deviation of $\sim$2.4$\%$ is significantly larger than the
statistical errors of the individual data bins.  With no adjustments,
the formal reduced $\chi^2$ would be much greater than one (4.6 in this
case).  This is, in general, an expected result since the observational
uncertainties from which $\chi^2$ is computed include statistical
errors only and are thus certainly underestimates.  In addition, it is
unlikely that the absolute photometric calibration of the data, the
extinction curve representation, or the atmosphere models are perfect
representations of reality. 

The rationale for adjusting the fitting weights is to yield more
realistic estimates of the parameter uncertainties, which scale as
$1/\sqrt{\chi^2}$.  The adjustment of the weights could be accomplished
in a number of ways, most simply by either scaling upward the
statistical errors of each bin by a single factor (2.1 in this case) or
by quadratically combining the statistical errors with an overall
uncertainty represented as a fraction of the local binned flux (2\% in
this case).  For HV~982 we adopted the latter technique, although both
yield virtually identical results (which are indistinguishable from the
no-adjustment case).

Note that this procedure is not rigorously justifiable, since it
implies that the discrepancies between the model and the data are due
entirely to underestimated observational errors.  However, the
resultant parameter uncertainties do appear reasonable for those cases
when an external check is possible.  We have such comparisons for two
parameters:  (1) $T_{eff}$ --- As we will show in \S 6 below, the
values of T$_{eff}$, with their attendant uncertainties, are completely
consistent with expectations from stellar evolution calculations; (2)
$\log g$ --- The fitting procedure utilizes the surface gravities
determined from the binary analysis.  However $\log g$ can also be
determined directly from the energy distributions.  When we fit the
HV~982 spectrum, constraining only the difference in $\log g$ between
the two components, we find for the primary star a value of $\log g =
3.83\pm0.07$, which compares very well with the observed value of
$3.78\pm0.03$ from the binary analysis.  We conclude that the adopted
procedure yields reasonable estimates of the true uncertainties of the
parameters determined from the energy distribution analysis.

The values of most of the parameters derived above will be discussed in
the sections below.  Here we comment briefly only on the results for
the STIS offsets and E(B-V).
  
The correction factors of 10.4\% and 6.4\% required to rectify the STIS
G430L and G750L spectra, respectively, are extremely well-determined
and appear surprising large.  Nevertheless, they are consistent with
the STIS calibration goals (of $\pm$10\%) for absolute photometry with
the CCD cameras.  The discrepancies may arise from several factors,
including the absolute photometric calibration, instrument stability,
and possible light loss in the 0.5\arcsec-wide slit.  The relative roles of
these various effects are uncertain at this point.  We have two other
LMC binary systems with similar sets of STIS and FOS spectra (EROS~1044
and HV~5936).  In all three cases, the STIS fluxes are below the FOS
level, with the G430L data always being the most discrepant.  Note
that, in the region of overlap between the FOS and STIS G430L, the {\it
shapes} of the spectra agree to within several percent, it is only the
general levels that are in discord.  We will examine the
cross-calibration of STIS and FOS using these and other data in a
future paper.

An accurate determination of ${\rm E(B-V)}$ is one of the most critical
factors in deriving accurate distances using this analysis --- and has
been one of the most problematic.  In this paper, the determination of
$\rm{E(B-V) = 0.086}$ is straightforward, highly precise, and
unambiguous because we have an ideal dataset consisting of
spectrophotometry spanning the entire wavelength range over which
$\rm{E(B-V)}$ is defined.  In an earlier version of this work
(Fitzpatrick et al. 2000), however, we performed the analysis utilizing
only FOS data (due to the lack of reliable optical photometry for
HV~982) and found a much different value of $\rm{E(B-V) \simeq 0.17}$.
It is clear in retrospect that the FOS data themselves, which truncate
at 4790 \AA, do not extend far enough into the optical region to allow
an accurate estimate of $\rm{E(B-V)}$.  With those data alone, ${\rm
E(B-V)}$ is wholly determined by only several hundred \AA\/ of spectrum
at the very end of the FOS G400H camera and is highly subject to any
systematic errors in the FOS photometric calibration or in the shape of
the adopted extinction curve in this small spectral region.  In fact,
we see a strong systematic effect in the analysis of three binaries for
which STIS data are now available --- the FOS data alone always yield
higher estimates of $\rm{E(B-V)}$ than when STIS spectra (or reliable
optical photometry) are incorporated in the analysis.  The moral of the
story is that a good determination of $\rm{E(B-V)}$ requires data which
span the wavelength range of the Johnson $V$ and $B$ filters, i.e., the
wavelength range over which $\rm{E(B-V)}$ is defined.
 
\section{The Interstellar Medium Towards HV~982}

Our analysis provides some general information regarding interstellar
gas and dust along the HV~982 sightline and also yields some insight
into the relative location of the star within the LMC.  This latter
point will prove important in interpreting the results from our
ensemble of LMC binaries.

As noted in \S3.3.2 above, we find a total H I absorption column
density of ${\rm N(H~I) = 1.55\times 10^{21} cm^{-2}}$ toward HV~982.
This consists of an assumed Galactic foreground contribution at 0 km
s$^{-1}$ of $5.5\times10^{20}$ cm$^{-2}$ (see, e.g., Schwering \&
Israel 1991) and an LMC contribution at 260 km s$^{-1}$ of
$1.0\times10^{21}$ cm$^{-2}$ determined by us from the strength of the
interstellar Ly$\alpha$ absorption line in the FOS G130H spectrum.  A
conservative estimate of the LMC column density uncertainty is
$\sim$$\pm$20\%.

\begin{figure*}
\plotone{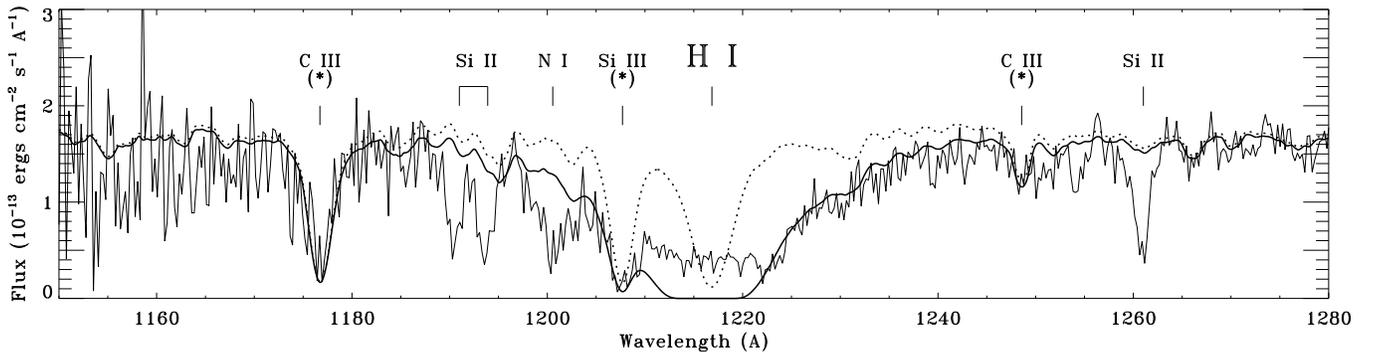}
\caption[f6.eps]{Derivation of the interstellar H I column
density towards HV~982.  The FOS data centered on the H~I Ly$\alpha$
line at 1215.7 \AA\/ are shown (thin solid line).  Prominent stellar
features (denoted with an asterisk) and interstellar features are
labeled.  The dotted line represents a synthetic spectrum of the HV~982
system, constructed by combining two individual velocity-shifted
spectra. The individual spectra were computed using Ivan Hubeny's {\it
SYNSPEC} spectral synthesis program with Kurucz {\it ATLAS9} atmosphere
models of the appropriate stellar parameters as inputs.  The solid
curve shows the synthetic spectrum convolved with an interstellar H~I
Ly$\alpha$ line computed with a Galactic foreground component of ${\rm
N(H~I) = 5.5\times10^{20} cm^{-2}}$ at 0 km s$^{-1}$ and a LMC
component of ${\rm N(H~I) = 1.0\times10^{21} cm^{-2}}$ at 260 km
s$^{-1}$ (see text in \S 4). \notetoeditor{THIS FIGURE IS INTENDED TO
SPAN TWO COLUMNS}
\label{figHI}}
\end{figure*}

The FOS data used to derive the H~I measurement are shown in Figure
\ref{figHI} where we plot a small section of the spectrum centered on
H~I Ly$\alpha$ with various stellar and interstellar lines labeled.
The spectral resolution and S/N of the data are not sufficient to
reveal the complex absorption profiles of the interstellar lines, which
span nearly 300 km s$^{-1}$ in velocity.  The dotted line shows a
synthetic spectrum of the HV~982 system, constructed from two
individual spectra, velocity shifted to match the stellar velocities at
the time of the FOS observations.  The individual spectra were computed
using {\it ATLAS9} model atmospheres of the appropriate physical
parameters and the spectral synthesis program {\it SYNSPEC}.
The thick solid line shows the synthetic spectrum convolved with an
interstellar Ly$\alpha$ profile computed using the velocities and
column densities noted in the paragraph above.  The adopted LMC column
density is that which produces the best agreement (based on visual
estimate) between the convolved spectrum and the data.  As can be seen
in Figure \ref{figHI}, the agreement is impressive, particularly in the
broad damping wings of Ly$\alpha$.  Note that the non-zero flux seen at
line center is due to geocoronal scattering of solar Ly$\alpha$
photons.

The H~I 21-cm emission line survey of Rohlfs et al. (1984) reveals that
the {\it total} LMC H~I column density along the HV~982 line of sight
is about $1.1\times 10^{21}$ cm$^{-2}$, centered at 260 km s$^{-1}$
(this velocity was adopted for the Ly$\alpha$ analysis).  This result,
combined with the Ly$\alpha$ determination for H~I {\it in front of}
HV~982, clearly demonstrates that HV~982 is actually located behind
most, if not all, of the LMC H~I along its line of sight.  We will
return to this point in \S 8.

\begin{figure*}
\plotone{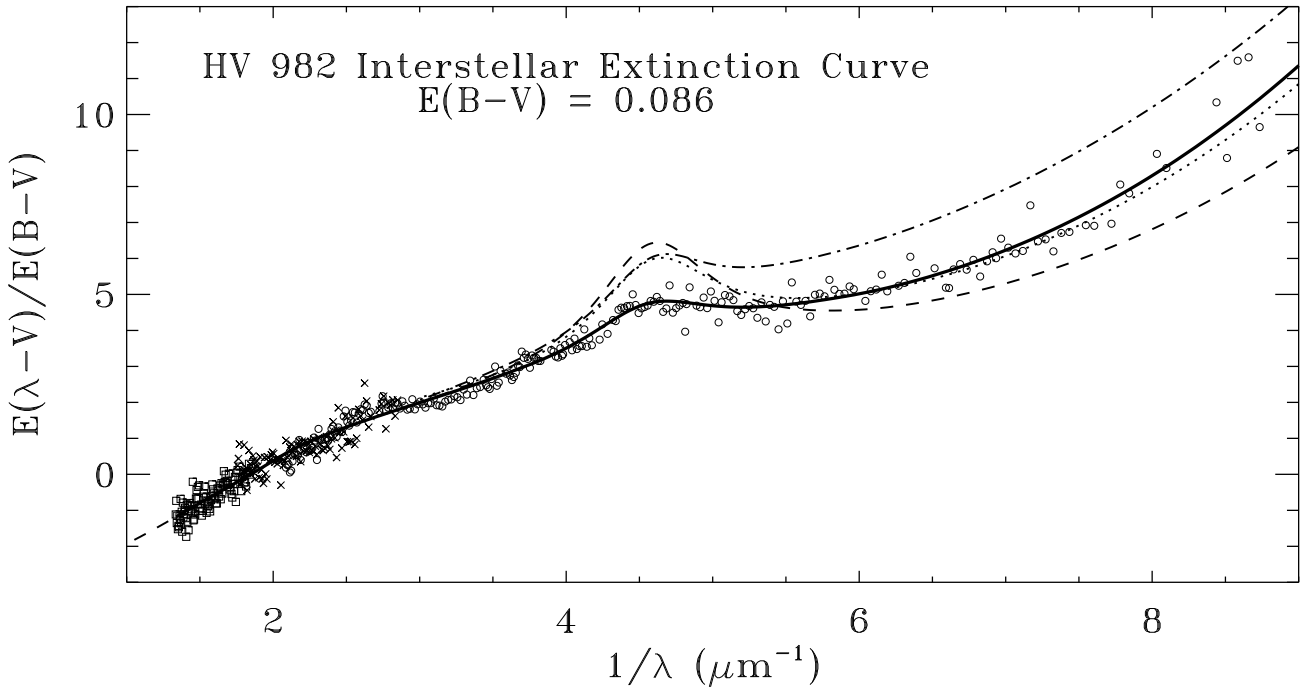}
\caption[f7.eps]{Normalized UV-through-optical interstellar
extinction curve for HV~982.  The thick solid line shows the
parametrized form of the extinction curve as determined by the SED
fitting procedure.  The recipe for constructing such a ``custom''
extinction curve is taken from F99 and the parameters defining it are
listed in Table \ref{tabEXT}.  Small symbols indicate the actual
normalized ratio of model fluxes to observed fluxes: circles, crosses,
and squares indicate FOS, STIS G430L, and STIS G750L data,
respectively.  Shown for comparison are the mean Milky Way extinction
curve for R = 3.1 from F99 (dashed line) and the mean LMC and 30
Doradus curves from Fitzpatrick (1986; dotted and dash-dotted lines,
respectively).  The HV~982 curve arises from dust in both the Milky Way
(${\rm E(B-V)_{MW} \simeq 0.06}$) and the LMC (${\rm E(B-V)_{LMC}
\simeq 0.027}$).  The main attributes of the curve, its high far-UV
level and extremely weak 2175 \AA\/ bump, are likely shared
characteristics of the extinction in both environments.
 \label{figEXT}}
\end{figure*}

The interstellar extinction curve determined for the HV~982 sightline
is shown in Figure \ref{figEXT}.  Small symbols indicate the normalized
ratio of model fluxes to observed fluxes, while the thick solid line
shows the parametrized representation of the extinction, which was
actually determined by the fitting process.  As noted in \S 3.3, the
recipe for constructing such a ``custom'' extinction curve is taken
from F99 and the parameters defining the curve are listed in Table
\ref{tabEXT}. 

Extinction-producing dust grains along the HV~982 line of sight lie in
both the Milky Way and the LMC.  The results of Oestreicher,
Gochermann, \& Schmidt-Kaler (1995) show that the Milky Way foreground
extinction in this direction corresponds to ${\rm E(B-V)_{MW}} \simeq 0.06$
mag.  Combined with our determination of the total value of
${\rm E(B-V)}$ (see Table \ref{tabPARMS}), this indicates an LMC
contribution of ${\rm E(B-V)_{LMC} \simeq 0.026}$.  The curve in Figure
\ref{figEXT} is thus clearly a composite, but weighted more heavily
towards the Galactic extinction component.  This prevents detailed
conclusions about either extinction component, although it is useful to
note that the width and position of the remarkably weak 2175 \AA\/
extinction bump are consistent with Galactic values (Fitzpatrick \&
Massa 1990).  Further, it is reasonable to conclude that the very weak
2175 \AA\/ bump is a feature of both the Milky Way and LMC extinction
components along this sightline.
 
Several of the stars in our LMC distance program have little or no
extinction beyond the Milky Way foreground contribution.  For these
stars we will be able to derive explicitly the shape of the Galactic
foreground extinction and perhaps use this result to ``deconvolve''
composite curves such as that for HV~982.

The ``gas-to-dust'' ratio for the LMC interstellar medium towards
HV~982 can be computed from the results above and is ${\rm N(H
I)/E(B-V)_{LMC} = 3.8 \times 10^{22} cm^{-2} mag^{-1}}$.  This value is
consistent with the range of values seen by Fitzpatrick (1986) for a
variety of LMC sightlines and is significantly higher than the mean
Milky Way value of ${\rm 4.8 \times 10^{21} cm^{-2} mag^{-1}}$ (Bohlin
et al. 1978), which may reflect the lower abundance of metals in the
LMC.

A final comment on extinction concerns the value of $R(V)$ (${\rm\equiv
A(V)/E(B-V)}$). In this analysis we adopt the mean Galactic value of $R(V)
= 3.1$ due, essentially, to a lack of any other option.  An individual
determination for a relatively lightly-reddened star like HV~982 would
require higher precision near-IR photometry than is currently available
(from {\it 2MASS}).  A correlation between $R(V)$ and the slope of
extinction in the UV (as parametrized by the fit coefficient {\it
$c_2$}) has been shown for a small sample of Galactic stars (F99), but
it cannot be assumed that such a correlation is applicable to a mixed
LMC/Galactic halo sightline.  The safest course for our analysis is to
adopt a value of $R(V)$ and then incorporate a reasonable estimate of its
uncertainty in the final error analysis.  As will be shown in \S 6, the
uncertainty in HV~982's distance due to $R(V)$ is only a small component
of the overall error budget.

\section{The Physical Properties of the HV~982 Stars}

The results of the analyses described above can be combined to provide
a detailed characterization of the physical properties of the HV~982
system.  We summarize these properties in Table \ref{tabSTAR}; notes to
the Table indicate how the individual stellar properties were derived
from the analysis.

It is important to realize that the results in Table \ref{tabSTAR} were
derived completely independently of any stellar evolution
considerations.  Thus, stellar structure and evolution models can be
used to provide a valuable check on the self-consistency of our
empirical results.  To test this consistency, we considered the
evolutionary models of Claret (1995, 1997) and Claret \& Gim\'enez
(1995, 1998) (altogether referred to as the CG models). These models
cover a wide range in both metallicity ($Z$) and initial helium
abundance ($Y$), incorporate the most modern input physics, and adopt a
value of 0.2~H$_{\rm p}$ as the convective overshooting parameter.

\begin{figure*}
\plotone{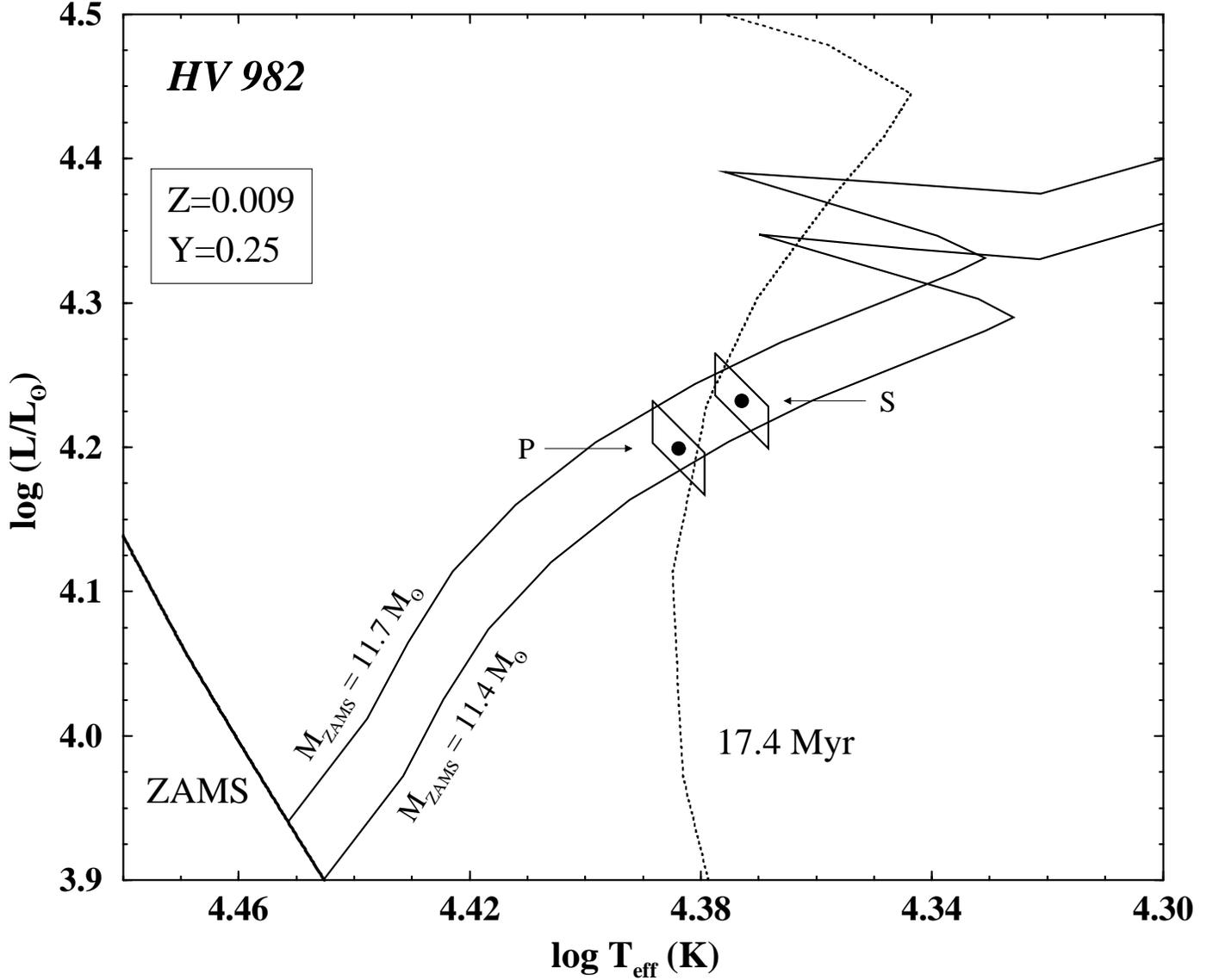}
\caption[f8.eps]{A comparison of the HV~982 results with stellar
evolution theory.  The positions of the the primary (P) and secondary
(S) components of HV~982 on the $\log L$ vs. $\log T_{\rm eff}$ diagram
are indicated by the filled circles.  The skewed rectangles represent
the 1$\sigma$ error boxes.  The position of the Zero Age Main Sequence
(ZAMS) is noted.  The two stellar evolution tracks shown (solid curves)
are not ``best fits.''  They correspond to the masses derived from the
binary analysis (which are $\sim0.1 M_{\sun}$ smaller than the original
ZAMS masses due to stellar wind mass loss) and the metallicity measured
from the UV/optical spectrophotometry.  Only the helium abundance Y has
been adjusted and the resultant value lies well within the expected
range.  The dotted line shows an isocrone corresponding to an age of
17.4 million years.  The source and properties of the evolution tracks
are discussed in \S4.
\label{figHRD}}
\end{figure*}

The locations of the HV~982 components in the $\log T_{\rm eff}$ vs.
$\log L$ diagram are shown in Figure \ref{figHRD}. The skewed
rectangular boxes indicate the 1$\sigma$ error locus (recall that
errors in $T_{\rm eff}$ and $L$ are correlated). If our results are
consistent with stellar evolution calculations, then the evolution
tracks corresponding to the masses and metallicity derived from the
analysis should pass through the error boxes in the $\log T_{\rm eff}$
vs. $\log L$ diagram. Indeed, this is the case. The thin solid lines
show the tracks corresponding to ZAMS masses of 11.7 and 11.4
$M_{\sun}$ and $Z = 0.009$ (based on the value of ${\rm [Fe/H]}$ resulting
from the spectrophotometry analysis). The models predict that such
stars should lose about 0.1 $M_{\sun}$ due to stellar winds by the time
they reach the positions of the HV~982 stars, yielding the present-day
masses of 11.6 and 11.3 $M_{\sun}$.  Further, the two stars are
compatible with a single isochrone, corresponding to an age of 17.4 Myr
(dotted line in Figure \ref{figHRD}).  Note that the only adjustable
parameter in the model comparison is the initial helium abundance $Y$,
for which we find an optimal value of $Y=0.25\pm0.03$. This value is in
excellent agreement with expectations from empirical chemical
enrichment laws (see Ribas et al. 2000b).

An additional, independent test of the compatibility of our results and
stellar structure theory can be made because the HV~982 system has an
eccentric orbit and a well-determined apsidal motion rate
($\dot{\omega}$).  The value of ($\dot{\omega}$) can be found by
combining the individual results for $\omega$ determined from the light
curve and the radial velocity curve analyses (see Table
\ref{tabPARMS}).  The resultant apsidal motion rate is $\dot{\omega} =
2.09 \pm 0.17$~deg/yr.  This is marginally consistent with the value of
$1.76\pm0.06$~deg/yr from P98, based on eclipse timings, although the
error in P98's result is likely to be underestimated due to large
uncertainties in some of the earliest timings.

The expected value of $\dot{\omega}$ for a binary system can be
computed as the sum of a general relativity term (GR) and a classical
term (CL). The latter, which is the most important contribution in
close systems like HV~982, depends on the internal mass distributions
of the stars, which can be derived from stellar evolution models.  The
mass concentration parameters $k_2$ (i.e, the ratio of the central
density to mean density) for the appropriate CG models (i.e., those
shown in Fig. \ref{figHRD}) are $\log {k_2}_{\rm P}=-2.34\pm0.03$ and
$\log {k_2}_{\rm S}=-2.30\pm0.03$.  (The error bars reflect the
observational uncertainties in the stellar masses.) These values
contain a small correction for stellar rotation effects, according to
Claret \& Gim\'enez (1993).  Using the formulae of Claret \& Gim\'enez,
we compute a classical term of ${\dot{\omega}}_{\rm CL}=1.90$~deg/yr
and a general relativistic term of ${\dot{\omega}}_{\rm GR}=
0.10$~deg/yr, yielding a total theoretical apsidal motion rate of
$\dot{\omega} {\rm (th)}=2.00\pm0.21$~deg/yr.  this result is in
excellent agreement with the observed values, once again demonstrating
consistency between our results and stellar interior theory.

Note that the eccentric nature of HV~982's orbit is not surprising
since circularization is not expected to occur until a later
evolutionary phase, when the stars have expanded to nearly fill their
Roche lobes.

Final reality checks on our results can be obtained from the
disentangled spectra of the primary and secondary components of the
HV~982 system, discussed in \S 3.1.  First, and most simply, the
spectra of these stars are completely consistent with the parameters
(particularly $T_{eff}$ and $\log g$) derived in our analysis, as is
well-illustrated in Figure \ref{figSECON}.  Second, the values of $v
\sin i$ (see Table \ref{tabSTAR}) also provide a remarkable
confirmation of our results.  These were derived by fitting the
disentangled spectra with a grid of synthetic spectra computed with $v
\sin i$ values ranging from 20 to 160 km s$^{-1}$.  The uncertainties
in the best-fitting values of $v \sin i$ were estimated from the
scatter in the results when small sections of the 4000---5000 \AA\/
disentangled spectra were considered individually.  These measured $v
\sin i$ values can be compared with theoretical expectations since the
HV~982 stars are expected to have undergone ``pseudo-synchronization,''
in which the stellar rotational speeds are determined by the orbital
angular velocity at periastron and the individual stellar radii.  For a
binary such as HV~982, pseudo-synchronization should occur in only
$\sim2$ Myr (using the recipe of Claret, Gim\'{e}nez, \& Cunha 1995),
much less than the age of the system.  The pseudo-synchronized values
of $v \sin i$ for the primary and secondary stars are 89.5 km s$^{-1}$
and 97.7 km s$^{-1}$, respectively (as computed from the results in
Kopal 1978), in excellent agreement with the measurements.

\section{The Distances to HV~982 and HV~2274}
\subsection{HV~982}

Our analysis has shown that HV~982 is an extraordinarily
well-characterized system consisting of a pair of normal,
mildly-evolved, early-B stars.  The results are all internally
consistent and consistent with a host of external reality checks,
such as the expected LMC metallicity, $M_V$ calibrations of Galactic B
stars, stellar evolution calculations, and binary evolution
calculations. This detailed characterization and the unremarkable
nature of the HV~982 stars make this system ideal for the determination
of a precise distance.

As in Paper I, we derive the distance to the system simply by combining
results from the EB analysis --- which yields the absolute radius of
the primary star $R_P$ --- and from the spectrophotometry analysis ---
which yields the distance attenuation factor $(R_P/d)^2$.  The result,
shown in Table \ref{tabSTAR}, is $d_{HV~982} = 50.2\pm1.2$ kpc
corresponding to a distance modulus of $(V_0 - M_V)_{HV~982} =
18.50\pm0.06$.

The uncertainty in the HV~982 distance determination arises from three
independent sources:  (1) the internal measurement errors in $R_A$ and
$(R_A/d)^2$ given in Table \ref{tabPARMS}; (2) uncertainty in the
appropriate value of the extinction parameter $R(V)$; and (3)
uncertainty in the FOS flux scale zeropoint due to calibration errors
and instrument stability.  Straightforward propagation of errors shows
that these three factors yield individual uncertainties of $\pm0.93$
kpc, $\pm0.52$ kpc (assuming $\sigma R(V) = \pm0.3$), and $\pm0.64$ kpc
(assuming $\sigma f(FOS) = \pm2.5$\%), respectively.  The overall
1$\sigma$ uncertainty quoted above is the quadratic sum of these three
errors.

Note that the only ``adjustable'' factor in the analysis is the
extinction parameter $R(V)$, for which we have assumed the value 3.1.
The weak dependence of our result on this parameter is given by:  $(V_0
- M_V)_{HV~982} = 18.50 - 0.075 \times [R(V)-3.1]$.

\subsection{HV~2274 (Again)} 

As noted in \S 1, there is a ``mini-controversy'' over the distance to
HV~2274, the LMC EB system we analyzed in Paper I.  Close examination
of the results from the various groups involved (including our own
early result for HV~2274 reported by Guinan et al. 1998b), reveals that
the discord among the various results arises from the inclusion of 
ground-based optical photometry in the energy distribution portion of
the analysis.  The problem is twofold:  First, there are significant
differences among the available {\it UBV} measurements for HV~2274 ---
differences larger than the claimed errors (see table 2 of Nelson et
al. 2000). Second, there is no obvious or objective way to determine
how the various optical photometric indices should be weighted in the
SED analysis with respect to the spectrophotometry.  

The need to incorporate ground-based photometry in the HV~2274 analysis
is clear: ${\rm E(B-V)}$ cannot be determined reliably unless the
energy distribution data extend through the $V$ spectral region (see
the discussion at the end of \S 3.3.3) and the available FOS
spectrophotometry for HV~2274 truncate in the $B$ region at 4790 \AA.
The complications arise in determining which data to use and how to use
them.

Since Paper I, we have studied extensively the effects of optical
photometry on the SED analysis and have concluded that the best way to
analyze datasets such as that for HV~2274 is also the simplest way,
namely, utilize the FOS spectrophotometry and only the $V$ magnitude
and force the best-fitting model to agree exactly with $V$.  The
benefits of this are that it eliminates the need to determine
weighing factors for the optical photometry and, importantly, allows
the effect of the adopted $V$ magnitude on the resulting distance
estimate to be examined explicitly.  The exclusion of $B-V$ and $U-B$
from the analysis is not a loss since, even in the best of
circumstances (i.e, no observational errors), they are wholly
redundant.

We have re-run the energy distribution portion of the HV~2274 analysis
using only the FOS data described in Paper I and a value of $V = 14.16$
(Udalski et al. 1998; Watson et al. 1992), which we believe to be
well-determined.  The FOS data processing and the fitting procedure was
applied exactly as described here for HV~982, with the addition of the
$V$ magnitude constraint.  Synthetic $V$ photometry was performed on
the models as described by FM99.  The parameters determined in the
analysis are the same as for HV~982, namely,  $T_{\rm eff}^P$,
[m/H]$_{PS}$, $\mu_{PS}$, $(R_P/d)^2$, $E(B-V)$, and six $k(\lambda-V)$
parameters.  Before fitting the energy distribution, we corrected the
HV~2274 FOS spectrum for presence of a strong interstellar H I
Ly$\alpha$ feature.  From fitting the broad line profile (using the same
technique as described in \S4 above for HV~982), we find a LMC H I
absorption column density of ${\rm 6.0\times10^{20} cm^{-2}}$ towards
HV~2274.  The LMC H~I 21-cm emission column density in this same
direction is ${\rm 1.2\times10^{21} cm^{-2}}$ (Rohlfs et al. 1984).

The results of this reanalysis are actually nearly identical to those
reported in Paper I and none of the conclusions in Paper I or Ribas et
al. (2000a) regarding the stellar properties of HV~2274 and their
consistency with stellar evolution theory are altered.  This agreement
--- in hindsight --- is not surprising since in Paper I we gave high
weight to the $V$ magnitude of Udalski et al. 1998 and low weights to
the $U$ and $B$ data due to conflicting observational reports.  We
thus, by accident, approached what we now believe is the optimal way to
combine the FOS and photometric datasets.

The distance derived for HV~2274 from the radius of the primary star
($R_P = 9.84$ R$_{\odot}$) and the parameter $(R_P/d)^2 = 2.228 \times
10^{-23}$ is $d_{HV~2274} = 47.0\pm2.2$ kpc corresponding to a distance
modulus of $(V_0 - M_V)_{HV~2274} = 18.36\pm0.10$.  This distance is
slightly larger than the value 46.8 kpc reported in Paper I.  The
error analysis incorporated (1) the internal uncertainties in $R_P$ and
$(R_P/d)^2$, (2) an uncertainty of $\pm0.3$ in the adopted value of
$R(V) = 3.1$, (3) an uncertainty of $\pm2.5$\% in the FOS flux scale,
and (4) an uncertainty of $\pm0.015$ in $V$.  These are all independent
effects and were combined quadratically to yield the quoted 1$\sigma$
error in $d_{HV~2274}$ and $(V_0 - M_V)_{HV~2274}$.  This result is
larger than the error of $\pm1.6$ kpc quoted in Paper I, due to the
more realistic treatment given to the effects of uncertainty in $V$.

As in the case of HV~982, the only ``adjustable'' parameter in this
result is the assumed value $R(V) = 3.1$.  Its influence, and the
explicit effect of the $V$ magnitude, on the result can be expressed
as:  $(V_0 - M_V)_{HV~2274} = 18.36 - 0.12 \times [R(V)-3.1] + 3.2
\times [V-14.16]$. Note that ${\rm E(B-V)}$ is not an adjustable
parameter --- its value (0.12 mag) is fully determined by the
analysis.  The HV~2274 distance is more sensitive to the uncertainty in
$R(V)$ than for HV~982 because of HV~2274's larger reddening.

\section{The Distance to the LMC}
 
\begin{figure*}
\plotone{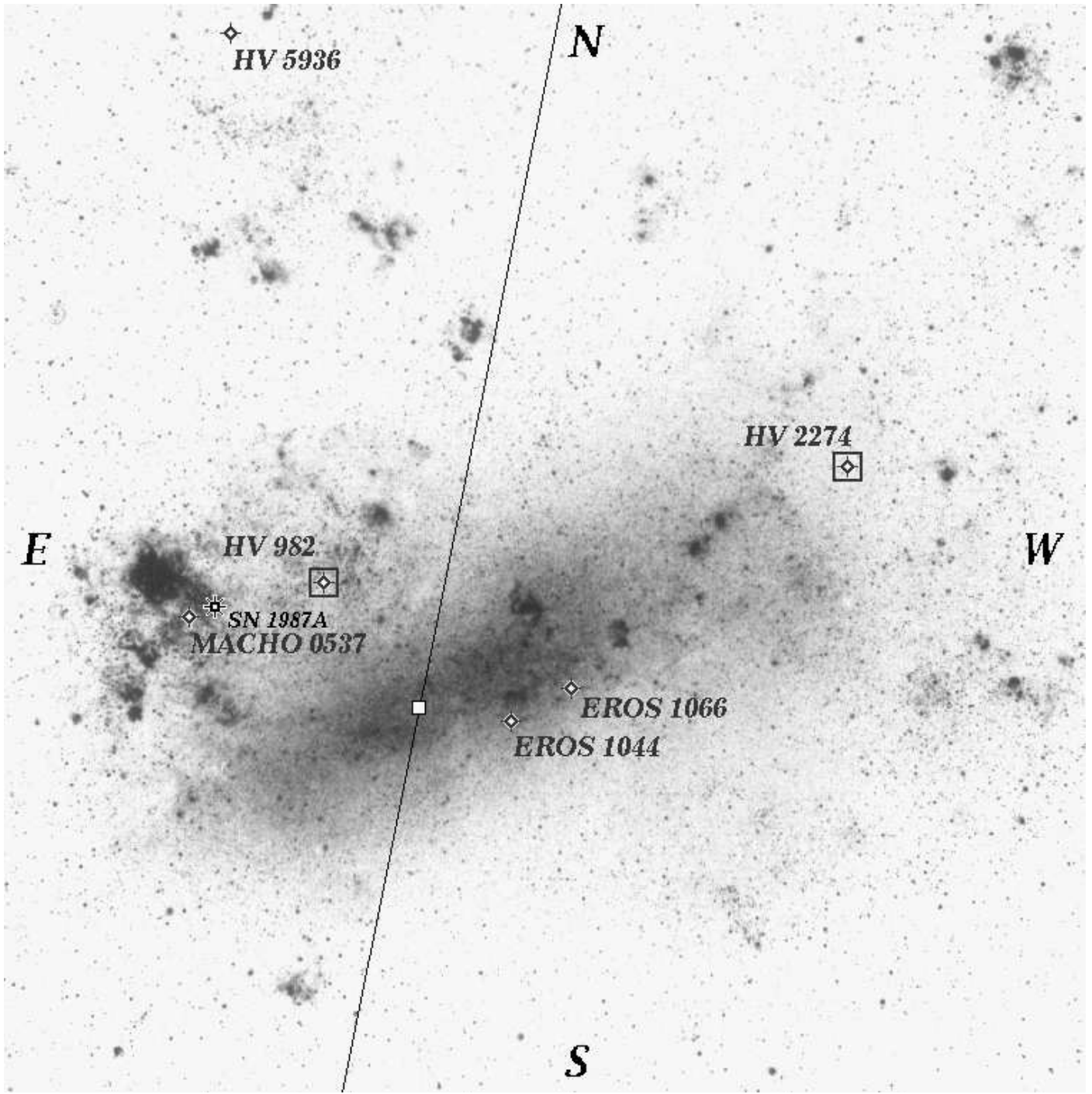}
\caption[f9.eps]{A photo of the Large Magellanic Cloud
indicating the locations of HV~982 (this paper), HV~2274 (Paper I), and
four targets of future analyses, HV~5936, EROS~1044, EROS~1066, and
MACHO 053648.7-691700 (labeled in the figure as MACHO 0537).  The
optical center of the LMC's bar according to Isserstedt 1975 is
indicated by the open box and the LMC's line of nodes, according to
Schmidt-Kaler \& Gochermann 1992, is shown by the solid line.  The
``nearside'' of the LMC is to the east of the line of nodes.  The
location of SN 1987A is also indicated. Photo reproduced by permission
of the Carnegie Institution of Washington.  \notetoeditor{THIS FIGURE IS INTENDED TO SPAN TWO COLUMNS}
\label{figLMC}}
\end{figure*}

Determining the distance to the LMC from the individual distances to
HV~982 and HV~2274 requires correcting for the spatial orientation of
the LMC's disk and the stars' apparent locations within it. We adopt as
a reference point the optical center of the LMC's bar, at ($\alpha$,
$\delta$)$_{1950}$ = ($5^h 24^m$, $-69\arcdeg\/ 47\arcmin$) according to
Isserstedt (1975), assume a disk inclination of $38\arcdeg$, and a
line-of-nodes position angle of $168\arcdeg$ (Schmidt-Kaler \&
Gochermann 1992).  Figure \ref{figLMC} shows a photo of the LMC with the
adopted center and the orientation of the line-of-nodes indicated by
the open box and solid line, respectively.  The ``near-side'' of the
LMC is eastward of the line-of-nodes.

The HV~982 system is located, in projection, relatively close to the
line-of-nodes (see Fig. \ref{figLMC}) and, given its measured distance,
should  --- {\it if it lies in the LMC's disk} --- be positioned about
450 pc in front of the bar center.  This would imply a distance to the
LMC bar center of 50.7 ($\pm1.2$) kpc, corresponding to a distance
modulus of 18.52 ($\pm0.06$) mag.

As discussed in Paper I, HV~2274's location (on the "far-side" of the
LMC; see Fig. \ref{figLMC}) places it --- {\it if it lies in the LMC's
disk} --- about 1100 pc beyond the bar center.  This result implies a
distance to the bar center of 45.9 ($\pm2.2$) kpc, corresponding to a
distance modulus of 18.31 ($\pm0.10$) mag.

These two estimates of the distance to the LMC bar differ by nearly 5
kpc.  (The magnitude of the discrepancy depends somewhat on the adopted
orientation of the LMC disk; however, this effect is only at the level
of a few hundred pc given the various estimates for the LMC's geometry;
see Westerlund 1997, page 30.) If the uncertainties in the two results
were uncorrelated, then this would amount to a $\sim$2$\sigma$
difference.  However, the uncertainties in the two results are not
completely uncorrelated.  For example, errors in the FOS flux zeropoint
would affect both analyses in the same way, as (probably) would errors
in the adopted value of $R(V)$ due to the similar lines of sight
through the Milky Way halo. In addition, the stars in the HV~982 and
HV~2274 systems bear close resemblance to each other.  Any small
systematic effects in the analyses would affect both results in a
similar way.  As a result, the discrepancy between the two measurements
of the LMC distance is likely to be somewhat larger than the formal
$\sim$2$\sigma$.

If these results represent marginally consistent, independent measures
of the same quantity ({\it the} LMC distance) then they imply a mean
distance of between 46 and 51 kpc.  (Including the EROS~1044 result
noted below, that mean is $\sim$48 kpc).  An alternate interpretation,
however, is that the HV~982 and HV~2274 results are discrepant because
the systems do not lie in a common disk (with the orientation usually
ascribed to the LMC), and therefore do not both constrain the distance
to this disk.  In this scenario, HV~2274 is associated with the disk
--- as indicated by the comparison of H~I emission and absorption
column densities in its direction (i.e., it lies behind $\sim50\%$ of
the H I in its direction) --- and HV~982 is located several kpc behind
the disk --- consistent with it's lying behind all the H~I in its
direction (see \S 4).  This scenario would indicate a significant depth
to the massive star distribution of the LMC and would have implications
for the value of the LMC as a calibrator of the cosmic distance scale.

The results for HV~982 and HV~2274, by themselves, do not persuasively
argue for the existence of a ``thick'' LMC.  Some support for this
hypothesis is available, however, from other data.  Specifically: 1)
the agreement between HV~982's distance and that of the nearby
SN~1987A ($51.4\pm1.2$ kpc, Panagia 1999); 2) the agreement between
the HV~2274 result and that for the EB system EROS~1044 (Ribas et al.
2001) which is also located in the LMC's bar and implies a LMC center
distance of $46.1\pm1.4$ kpc; and 3) the discrepancy between red clump
distance estimates for the 30 Doradus region ($52.2 \pm 2.1$ kpc;
Romaniello et al. 2000) and the LMC bar ($44.5 \pm 1.4$ kpc; Udalski
2000).  In general, the existence of significant line-of-sight
structure in the LMC would not be surprising, given its history of
gravitational interaction with the Milky Way (e.g., Weinberg 2000),
although this hypothesis may be difficult to reconcile with some
observational indications of strong regularity within the system (see,
e.g., the H~I synthesis maps of Kim et al. 1998).

Clearly, additional results are needed to determine the extent of
line-of-sight structure in the LMC and to derive a best estimate of the
distance to the system.  We have completed analysis of two more EB
systems, EROS~1044 (Ribas et al. 2001) and HV~5936 (in preparation),
and have begun work on two more, EROS~1066 and MACHO~0537. (See Figure
\ref{figLMC} for the locations.)  For all of these systems,
spectrophotometry spanning the range 1150 \AA\/ to 7500 \AA\/ has been
obtained and the quality of the individual distance determinations will
be comparable to that for HV~982 in this paper.  Within the next few
years we hope to expand the program to include about 20 systems.  Our
overall ensemble of targets, in addition to nailing down the distance
to the LMC, will provide a detailed probe of the structure and spatial
extent of this important galaxy.

\acknowledgements

This work was supported by NASA grants NAG5-7113, HST GO-06683, HST
GO-08691, and NSF/RUI AST-9315365.  We are grateful for the skilled
assistance of the CTIO support staff during our January and December
2000 observing runs.  E.F. acknowledges support from NASA ADP grant
NAG5-7117 to Villanova University and thanks Michael Oestreicher for
kindly making his LMC foreground extinction data available.  I.R.
acknowledges the Catalan Regional Government (CIRIT) for financial
support through a postdoctoral Fulbright fellowship.  We thank the
referee Mario Mateo for a variety of helpful suggestions and
comments.
 


\begin{deluxetable}{cccccc}
\tablewidth{0pc}
\tablecaption{Heliocentric Radial Velocity Measurements for HV~982}
\tablehead{
\colhead{HJD}        & 
\colhead{Orbital}    &  
\colhead{RV$_P$}     &
\colhead{RV$_S$}     &
\colhead{(O-C)$_P$}  &
\colhead{(O-C)$_S$}  \\
\colhead{($-$2400000)}  &
\colhead{Phase}         & 
\colhead{(km s$^{-1}$)} &
\colhead{(km s$^{-1}$)} &
\colhead{(km s$^{-1}$)} &
\colhead{(km s$^{-1}$)} }
\startdata
51558.6813 &   0.7203 &   446.2 &  135.6 &\phm{$-$}3.5 &  $-1.3$    \\
51558.7073 &   0.7252 &   444.8 &  137.3 &\phm{$-$}3.3 &  $-0.8$    \\
51560.6335 &   0.0862 &   185.1 &  383.7 &  $-1.5$     &  $-1.8$    \\
51560.6589 &   0.0910 &   182.5 &  389.2 &\phm{$-$}0.0 &  $-0.2$    \\
51560.7966 &   0.1168 &   158.0 &  409.1 &  $-2.7$     &  $-1.5$    \\
51561.7346 &   0.2926 &   107.8 &  462.0 &\phm{$-$}1.9 &  $-2.1$    \\
51561.7569 &   0.2968 &   107.3 &  463.8 &  $-0.8$     &\phm{$-$}1.8\\
51561.7823 &   0.3016 &   110.4 &  459.6 &  $-0.4$     &\phm{$-$}0.2\\
51563.7379 &   0.6681 &   448.3 &  128.7 &  $-1.7$     &  $-1.1$    \\
51563.7601 &   0.6723 &   450.2 &  128.8 &\phm{$-$}0.4 &  $-1.1$    \\
51897.5627 &   0.2381 &\phn97.7 &  472.8 &\phm{$-$}2.0 &  $-1.0$    \\
51897.5859 &   0.2425 &\phn98.1 &  473.7 &\phm{$-$}2.7 &  $-0.4$    \\
51899.5515 &   0.6109 &   439.7 &  138.1 &  $-1.6$     &  $-0.1$    \\
51899.5730 &   0.6149 &   443.4 &  135.9 &\phm{$-$}0.8 &  $-1.0$    \\
51900.5567 &   0.7993 &   409.5 &  168.2 &  $-1.9$     &\phm{$-$}0.9\\
51900.5764 &   0.8030 &   410.7 &  171.5 &\phm{$-$}1.3 &\phm{$-$}2.2\\
51902.5584 &   0.1745 &   119.2 &  452.7 &  $-0.2$     &\phm{$-$}2.0\\
51902.5782 &   0.1782 &   118.1 &  452.1 &\phm{$-$}0.9 &  $-0.7$    \\
\enddata
\label{tabRV}
\end{deluxetable}


\small
\begin{deluxetable}{lc}
\tablewidth{0pc}
\tablecaption{Results From Light Curve, Radial Velocity Curve, and Spectrophotometry Analyses}
\tablehead{\colhead{Parameter} & \colhead{Value}}
\startdata
\multicolumn{2}{c}{{\it Radial Velocity Curve Analysis}}                         \\
$\omega$ (2000.5) (deg)                               & $237.1\pm1.0$\phn\phn    \\
$K_{\rm P}$ (km~s$^{-1}$)                             & $177.7\pm3.2$\phn\phn    \\
$K_{\rm S}$ (km~s$^{-1}$)                             & $172.7\pm3.2$\phn\phn    \\
$q\equiv\frac{M_{\rm S}}{M_{\rm P}}$                  & $1.029\pm0.027$          \\
$\gamma$ (km~s$^{-1}$)                                & $287.8\pm2.5$\phn\phn    \\
$a$ (R$_{\odot}$)                                     & $36.5\pm0.5$\phn         \\
\multicolumn{2}{c}{{\it Light Curve Analysis}}    \\
Period (days)                                         & $5.335220\pm0.000003$ \\
Eccentricity                                          & $0.156\pm0.005$       \\
Inclination (deg)                                     & $89.3\pm0.7$\phn      \\
$\omega$ (1994.0) (deg)                               & $223.5\pm0.5$\phn\phn \\
$T_{\rm eff}^{\rm S}/T_{\rm eff}^{\rm P}$ & $0.975\pm0.005$       \\
$[L_{\rm S}/L_{\rm P}]_{V,I}$   		      & $1.14\pm0.02$         \\
$r_{\rm P}$\tablenotemark{a}                          & $0.1965\pm0.0020$     \\
$r_{\rm S}$\tablenotemark{a}                          & $0.2146\pm0.0018$     \\
$\Omega_P$\tablenotemark{b}                                           & $6.357\pm0.038$          \\
$\Omega_S$\tablenotemark{b}                                            & $6.023\pm0.029$          \\  
\multicolumn{2}{c}{\it Energy Distribution Analysis}                             \\
$T_{\rm eff}^{\rm P}$ (K)                             &  $24200\pm250$\phn\phn   \\
${\rm[m/H]}_{\rm PS}$                                 &  $-0.45\pm0.05$\phm{$-$} \\
$\mu_{\rm PS}$ (km s$^{-1}$)                          &  0                       \\      
E(B$-$V) (mag)                                        &  $0.086\pm0.005$         \\
$\log (R_P/d)^2$                                      &  $1.038\pm0.016 \times 10^{-23}$  \\
\tablenotetext{a}{Fractional stellar radius $r \equiv R/a$, where 
$R$ is the stellar ``volume radius'' and $a$ is the orbital semi-major axis.}
\tablenotetext{b}{Stellar equipotential surfaces.}
\enddata
\normalsize
\label{tabPARMS}
\end{deluxetable}
\normalsize


\begin{deluxetable}{ccc}
\tablewidth{0pc}
\tablecaption{Offsets Applied to HST/STIS Observations}
\tablehead{
\colhead{HST Dataset}  & 
\colhead{STIS Grating} & 
\colhead{Offset}       \\
\colhead{Name}         & 
\colhead{ }            & 
\colhead{(${\rm FOS - STIS}$)} }
\startdata
O665A8030  & G430L &  +$10.4\pm0.5\%$  \\ 
O665A8040  & G750L & \phn+$6.4\pm0.7$\%\\ 
\enddata
\label{tabSTIS}
\end{deluxetable}


\begin{deluxetable}{clc}
\tablewidth{0pc}
\tablecaption{Extinction Curve Parameters for HV~982}
\tablehead{
\colhead{Parameter}   & 
\colhead{Description} & 
\colhead{Value}        }
\startdata
$x_0$       & UV bump centroid        & $4.57\pm0.03$ ${\rm\mu m^{-1}}$\\
$\gamma$    & UV bump FWHM            & $1.07\pm0.14$ ${\rm\mu m^{-1}}$\\
$c_1$       & linear offset           & $-0.78\pm0.24$                 \\
$c_2$       & linear slope            & $\phm{-}0.93\pm 0.06$          \\
$c_3$       & UV bump strength        & $\phm{-}1.45\pm 0.42$          \\
$c_4$       & FUV curvature           & $\phm{-}0.54\pm 0.07$          \\   
$R(V)$         & ${\rm A(V)/E(B-V)}    $  & 3.1 (assumed)                  \\
\enddata
\tablecomments{The extinction curve parametrization scheme is based on the work of Fitzpatrick \& Massa 1990 and the complete UV-through-IR curve is constructed following the recipe of Fitzpatrick 1999.}
\label{tabEXT}
\end{deluxetable}

 
\begin{deluxetable}{lcc}
\tabletypesize{\scriptsize}
\tablewidth{0pc}
\tablecaption{Physical Properties of the HV~982 System}
\tablehead{
\colhead{Property}     & 
\colhead{Primary}      &  
\colhead{Secondary}    \\
\colhead{}             &
\colhead{Star}         &  
\colhead{Star}          } 
\startdata
Spectral Type\tablenotemark{a}       & B1 V-IV                 & B1 V-IV                \\
$V$\tablenotemark{b} (mag)           & 15.46                   & 15.32                  \\
Mass\tablenotemark{c} (M$_{\sun}$)   & $11.3\pm0.5$\phn        & $11.6\pm0.5$\phn       \\
Radius\tablenotemark{d} (R$_{\sun}$) & $7.17\pm0.12$           & $7.83\pm0.13$          \\
$\log g$\tablenotemark{e} (cgs)      & $3.780\pm0.023$         & $3.716\pm0.023$        \\
T$_{eff}$\tablenotemark{f} (K)       & $24200\pm250$\phn\phn   & $23600\pm250$\phn\phn  \\
$\log (L/L_{\sun})$\tablenotemark{g} & $4.20\pm0.02$           & $4.23\pm0.02$          \\
$[$Fe/H$]$\tablenotemark{h}           & $-0.33\pm0.05$\phm{$-$} & $-0.33\pm0.05$\phm{$-$}\\
$v \sin i$\tablenotemark{i} (km s$^{-1}$) & $85\pm5$\phn   & $106\pm11$\phn  \\
$M_{bol}$\tablenotemark{j} (mag)     & $-5.75\pm0.05$\phm{$-$} & $-5.83\pm0.05$\phm{$-$}\\
$M_{V}$\tablenotemark{k} (mag)       & $-3.44$                 & $-3.52$                \\
Age\tablenotemark{l} (Myr)           & \multicolumn{2}{c}{$17.4$}  \\
$d_{HV982}$\tablenotemark{m} (kpc)   & \multicolumn{2}{c}{$50.2\pm1.2$}  \\
\tablenotetext{a}{Estimated from $T_{\rm eff}$ and $\log g$} 
\tablenotetext{b}{From synthetic photometry of best-fitting model to the HV~982 system.  Combining the magnitudes yields $V_{HV982} = 14.64$}
\tablenotetext{c}{From the mass ratio $q$ and the application of Kepler's Third Law.}
\tablenotetext{d}{Computed from the relative radii $r_P$ and $r_S$ and the orbital semimajor axis $a$.}
\tablenotetext{e}{Computed from $g = G M / R^2$}
\tablenotetext{f}{Direct result of the spectrophotometry analysis and photometrically-determined temperature ratio.}
\tablenotetext{g}{Computed from $L = 4 \pi R^2 \sigma T^4_{eff}$} 
\tablenotetext{h}{Direct result of the spectrophotometry analysis, adjusted by +0.12 dex to account for the overabundance of Fe in the Kurucz {\it ATLAS9} opacities.  See FM99.}
\tablenotetext{i}{$v \sin i$ measured from the ``disentangled spectra'' of the two components as described in the text in \S 5.}
\tablenotetext{j}{Computed from $\log (L/L_{\sun})$ and $M_{bol\sun} = 4.75$} 
\tablenotetext{k}{Computed from $M_{bol}$ and a bolometric correction of $BC = -2.31$ taken from Flower 1996 for $T = 24000$ K.  Note that this result is consistent with expectations for mildly evolved early-B stars.} 
\tablenotetext{l}{From the best-fitting isochrone to the data shown in Fig. \ref{figHRD}.} 
\tablenotetext{m}{Using $(R_P/d)^2$ from the spectrophotometry analysis and $R_P$ from the light curve and radial velocity curve analyses. See \S 6.} 
\enddata
\label{tabSTAR}
\end{deluxetable}
\clearpage


\newpage
\newpage

\end{document}